\documentclass[11pt]{article}
\pdfoutput=1

\usepackage{jheppub}
\usepackage{url,comment}
\usepackage{times}
\usepackage{latexsym}
\usepackage{graphicx, graphics, hyperref, amsmath, amssymb, slashed, color, bbm,amsthm, array}
 \usepackage{subfigure}
 
\newcommand{\nc}{\newcommand}

\nc{\beq}{\begin{equation}}
\nc{\eeq}{\end{equation}}
\nc{\barray}{\begin{eqnarray}}
\nc{\earray}{\end{eqnarray}}
\nc{\barrayn}{\begin{eqnarray*}}
\nc{\earrayn}{\end{eqnarray*}}
\nc{\bcenter}{\begin{center}}
\nc{\ecenter}{\end{center}}
\nc{\mc}{\mathcal}
\nc{\er}[1]{(\ref{eq:#1})}
\nc{\onehalf}{\frac{1}{2}} 
\nc{\partialbar}{\bar{\partial}}
\nc{\psit}{\widetilde{\psi}}
\nc{\Tr}{\mbox{Tr}}
\nc{\hc}{\mbox{H.c.}}
\nc{\ev}{\;\mathrm{eV}}
\nc{\mev}{\;\mathrm{MeV}}
\nc{\gev}{\;\mathrm{GeV}}
\nc{\tev}{\;\mathrm{TeV}}

\def\chii0{\chi_i^0}
\def\chij0{\chi_j^0}

\newcommand{\met}{E\!\!\!/_T}
\newcommand{\gsim}{\lower.7ex\hbox{$\;\stackrel{\textstyle>}{\sim}\;$}}
\newcommand{\lsim}{\lower.7ex\hbox{$\;\stackrel{\textstyle<}{\sim}\;$}}
\nc{\ttbar}{t\bar t}

\def\iab{{\ \rm ab}^{-1}}

\newcommand{\fref}[1]{Fig.~\ref{f.#1}}
\newcommand{\eref}[1]{Eq.~(\ref{e.#1})}

\newcommand{\aref}[1]{Appendix~\ref{a.#1}}
\newcommand{\sref}[1]{Section~\ref{s.#1}}
\newcommand{\ssref}[1]{Section~\ref{ss.#1}}
\newcommand{\sssref}[1]{Section~\ref{sss.#1}}
\newcommand{\cref}[1]{Chapter~\ref{c.#1}}

\newcommand{\MSbar}{$\overline{\mathrm{MS}}$~}

\newcommand{\co}[1]{\mathcal{O}(#1)}

\graphicspath{{plots/}}

\title{Testing Electroweak Baryogenesis with Future Colliders}

\author{David Curtin,}
\author{Patrick Meade,}
\author{Chiu-Tien Yu}

\affiliation{C. N. Yang Institute for Theoretical Physics, Stony Brook University, Stony Brook, NY 11794\\}

\emailAdd{\\david.curtin@stonybrook.edu}
\emailAdd{\\patrick.meade@stonybrook.edu}
\emailAdd{\\chiu-tien.yu@stonybrook.edu}

\abstract{Electroweak Baryogenesis (EWBG) is a compelling scenario for explaining the matter-antimatter asymmetry in the universe. Its connection to the electroweak phase transition makes it inherently testable. However, completely excluding this scenario can seem difficult in practice, due to the sheer number of proposed models. We investigate the possibility of postulating a ``no-lose'' theorem for testing EWBG in future  $e^+e^-$ or hadron colliders. As a first step we focus on a factorized picture of EWBG which separates the sources of a stronger phase transition from those that provide new sources of CP violation. We then construct a ``nightmare scenario'' that generates a strong first-order phase transition as required by EWBG, but is very difficult to test experimentally. We show that a 100 TeV hadron collider is both necessary and possibly sufficient for testing the parameter space of the nightmare scenario that is consistent with EWBG.
}

\begin{document}

\begin{flushright}
\small{YITP-SB-14-33}
\end{flushright}

\maketitle

\section{Introduction}\label{s.intro}

The origin of the matter-antimatter asymmetry of our observed universe remains one of the most important unsolved mysteries in particle physics.  This is not for a lack of compelling theoretical ideas, but rather due to the lack of compelling experimental evidence for any of those ideas.  
A hypothetical physical process in the early universe which generates the observed asymmetry between baryons and anti-baryons is called baryogenesis. This requires satisfying the three Sakharov conditions \cite{Sakharov:1967dj}, which in many theories is achieved by introducing new GUT or high-scale physics that is not directly accessible at collider experiments. By construction, these theories are difficult or impossible to test unambiguously. 

In contrast, the Standard Model (SM) itself contains all the necessary ingredients to realize the mechanism of \emph{Electroweak Baryogenesis} (EWBG)~\cite{Kuzmin:1985mm, Klinkhamer:1984di, Arnold:1987mh, Arnold:1987zg, Khlebnikov:1988sr}. Unfortunately, the actual parameters of the SM do not satisfy the Sakharov conditions, and thus EWBG also requires the introduction of new physics beyond the SM (BSM).

Even so, EWBG stands out from other baryogenesis scenarios simply because it occurs at or near the electroweak (EW) scale. The basic mechanism proceeds as follows (see \cite{
Cline:2006ts, Trodden:1998ym, Riotto:1998bt, Riotto:1999yt, Quiros:1999jp, Morrissey:2012db} for reviews). In the very early universe, interactions with the plasma stabilize the higgs field at the origin. As the universe cools down, the higgs undergoes a phase transition to a nonzero vacuum expectation value (VEV) when the temperature is in the neighborhood of the weak scale  $T \sim \mathcal{O}(100 \gev)$. If this phase transition is sufficiently first-order, $CP$-violating interactions of the plasma with the expanding bubble wall of true vacuum can generate a chiral excess in front of the wall, which is then converted to baryons by electroweak sphalerons. Given a strong enough phase transition, a sufficient portion of the generated baryon asymmetry survives inside the bubble of true vacuum once the bubble wall moves past. Both the strong phase transition and large $CP$-violation require new physics, which has to be active near the EW scale. This makes EWBG, in principle, \emph{fully testable} at collider experiments. 

The difficulty in testing EWBG arises from the multitude of proposed models \cite{
Cline:2006ts, Trodden:1998ym, Riotto:1998bt, Riotto:1999yt, Quiros:1999jp, Morrissey:2012db}, and a priori one would need to investigate the {\em entire} theory space of EWBG to determine the necessary reach of a future collider. Instead, we propose a systematic approach in which we closely examine the requirements that new physics must satisfy for successful EWBG, and then determine if there is an axis along which experimental testability becomes more difficult. We then look only at models in this most difficult regime. 

We set as our axes the two basic BSM requirements for successful EWBG: (i) a modification of the higgs potential at high temperatures to make the phase transition more first-order than in the SM, and (ii) some new form of $CP$-violation. In a particular model, there are different testable consequences along each of these axes. For example in the MSSM, the stronger phase transition requires particular spectra with light stops~\cite{
Huet:1995sh, Carena:1996wj, Laine:1998qk, Laine:2000xu, Espinosa:1996qw, Carena:1997ki, Huber:2001xf, Cline:2000kb, Carena:2002ss, Lee:2004we, Cirigliano:2009yd, Carena:2008rt, Carena:2008vj, Quiros:1999tx, Delepine:1996vn}.  Light stops in turn are easily testable both through direct searches and indirect properties of the higgs boson.  In fact, one can exclude EWBG in the MSSM using early higgs data without relying on direct searches by correlating the various different higgs production modes and decays~\cite{Curtin:2012aa, Cohen:2011ap, Cohen:2012zza}.  Studying the sources of CP violation also provides experimental tests; for instance, Electric Dipole Moments (EDMs) provide stringent tests of baryogenesis in the MSSM~\cite{Cirigliano:2009yd,Curtin:2012aa}.

In principle, for any given model we can look for the experimental consequences along both the phase transition axis and the CP violation axis.  However, the detailed calculation of the generated baryon asymmetry that relies on new sources of CP violation is extremely complex and subject to large theoretical uncertainties, see e.g. \cite{
Huber:2001xf, Cline:2000kb, Moore:2000wx, John:2000zq, John:2000is, Megevand:2009gh, Moreno:1998bq, Riotto:1998zb, Lee:2004we, Cirigliano:2006wh, Chung:2008aya, Carena:2002ss, Li:2008ez, Cline:1997vk, Cline:2000nw, Konstandin:2004gy, Konstandin:2003dx, Kozaczuk:2012xv}.  In contrast, determining if there is a strong first-order phase transition is much more tractable and can be factorized from the full problem. Therefore, to determine a {\em minimum} criterion for testing EWBG, we first look solely at the phase transition requirement. Although this does not test the full mechanism of EWBG, ruling out the possibility of a first-order phase transition is sufficient to rule out the possibility of EWBG. 
We do not, at the present time, investigate the correlation between BSM physics responsible for the sources of CP violation and the phase transition. This will only become necessary if there is a part of theory parameter space that is not testable through the phase transition requirement alone.  

A strong first-order electroweak phase transition is characterized by the presence of a barrier in the effective thermal higgs potential that separates degenerate minima $h = 0$ and $h = v_c$ at some temperature $T_c$, while satisfying the \emph{Baryon-Number Preservation Criterion}  for $m_h = 125 \gev$,
\begin{equation}
\label{e.bnpc}
\frac{v_c}{T_c} > 0.6 - 1.6.
\end{equation}
The right-hand side is conventionally taken to be 1.0, but we consider the shown numerical range to reflect unknown details of the baryon number generation mechanism during the phase transition \cite{Patel:2011th}.  There are a number of possible ways to achieve this phase transition by introducing new particles which couple to the higgs, modifying its potential by a variety of mechanisms (see~\cite{Chung:2012vg} for a categorization of phase transitions and their correlation to higgs observables). Therefore, moving down the axis of difficulty in testing sources of a strong phase transition is relatively straightforward. The most-hidden particles that can increase the strength of the EW phase transition are SM singlet scalars.  SM singlets that couple to the higgs to achieve EWBG have been studied in great detail, both by themselves~\cite{Profumo:2007wc, Ashoorioon:2009nf, Damgaard:2013kva,  Barger:2007im, Espinosa:2011ax, Noble:2007kk, Cline:2012hg, Cline:2013gha, Alanne:2014bra, Espinosa:2007qk, Profumo:2014opa, Fuyuto:2014yia,  Fairbairn:2013uta} and in the context of supersymmetry \cite{Pietroni:1992in, Davies:1996qn, Huber:2006wf, Menon:2004wv, Huber:2006ma, Huang:2014ifa, Kozaczuk:2014kva}. 
In this paper we investigate the maximally hidden singlet scalar model,  find where a strong phase transition can occur, and then correlate this with the reach of experimental probes. 

The basic setup for this ``nightmare scenario'' is as follows. We introduce a real singlet field that couples to the higgs and has a $\mathbb Z_2$ symmetry to forbid higgs-singlet mixing~\cite{Barger:2007im, Noble:2007kk, Espinosa:2011ax, Cline:2012hg, Cline:2013gha, Alanne:2014bra, Espinosa:2007qk}. This rules out electroweak precision tests and higgs coupling modifications as experimental probes. We then set $m_S > m_h/2$ to avoid modified higgs decays, in particular an exotic higgs decay mode which would be relatively easy to discover at future colliders (see \cite{Curtin:2013fra} for a review).  

This nightmare scenario, while difficult to test, still has a number of potential experimental signatures. For instance, colliders can probe the direct production of the singlet states, as well as shifts in the triple higgs couplings and $Zh$ cross section.  Furthermore, the presence of the $\mathbb Z_2$ symmetry has implications for dark matter searches.

One could, in principle, make the above setup even more difficult to discover by including extra singlets that decrease some of the experimental signatures while leaving the phase transition intact. However, as we show in this paper, excluding even this basic nightmare scenario \emph{requires} at least a 100 TeV hadron collider, such as the proposed SPPC/FCC. A higgs factory like  CEPC, ILC, or TLEP has some sensitivity but is not sufficient, based on existing studies for precision measurements of higgs couplings. 
Remarkably, the fact that this scenario {\em is} testable at the SPPC/FCC  demonstrates that it may be possible to postulate a ``no-lose" theorem for EWBG with future colliders.

Our paper is organized as follows. In \sref{scenario}, we define the $\mathbb{Z}_2$ symmetric singlet scalar model and the two-dimensional parameter plane that illustrates its entire phenomenology. \sref{ewpt} contains our analyses of the one-step and two-step phase transitions which enable EWBG in this model.  Sections~\ref{s.collider} and \ref{s.indirect} examine direct and indirect signatures of the singlet scalar at colliders, and show how the discovery potential overlaps with the EWBG-favored regions of parameter space. We consider cosmological constraints on the singlet in \sref{dm} and show that, under certain assumptions, the entire parameter space can be excluded by future direct detection experiments. Renormalization group (RG) evolution and the implications of strong couplings are discussed in \sref{strongcouplings}. We summarize our findings and discuss implications in \sref{discussion}.

\section{A ``Nightmare Scenario'' for a Strong Electroweak Phase Transition}\label{s.scenario}

 Our putative nightmare scenario is constructed to hide the effects of a strong first-order phase transition, as discussed in Section~\ref{s.intro}. 
 
\subsection{Model Definition}
We define our model by the following most general renormalizable tree-level higgs potential for the SM higgs and a single real scalar:
\begin{equation}
V_0 = - \mu^2 |H|^2 + \lambda |H|^4 + \frac{1}{2} \mu_S ^2 S^2 + \lambda_{HS} |H|^2 S^2 + \frac{1}{4} \lambda_S S^4.
\end{equation}
After substituting $H = (G^+, (h+ i G^0)/\sqrt{2})$ and focusing on the field $h$ which becomes the SM higgs after acquiring a VEV\footnote{For simplicity, we use $h$ for the neutral real component of $H$ as well as the SM higgs.}, this becomes
\begin{equation}
\label{e.V0}
V_0 = - \frac{1}{2} \mu^2 h^2 + \frac{1}{4} \lambda h^4 + \frac{1}{2} \mu_S^2 S^2 + \frac{1}{2} \lambda_{HS} h^2 S^2 + \frac{1}{4} \lambda_S S^4.
\end{equation}
This scenario of adding a singlet with a $\mathbb{Z}_2$ symmetry to the SM has been well-studied in a variety of different contexts~\cite{Barger:2007im, Noble:2007kk, Espinosa:2011ax, Cline:2012hg, Cline:2013gha, Alanne:2014bra, Espinosa:2007qk}. In this work, we focus on adding one real singlet with a mass larger than $m_h/2$ to avoid exotic higgs decays, and an unbroken $\mathbb{Z}_2$ symmetry under which $S \to -S$ to avoid singlet-higgs mixing.  In our choice of parametrization, the higgs acquires a VEV $\langle h \rangle = v  = \mu/\sqrt{\lambda} \approx 246 \gev$ and a mass at tree-level $m_h = \sqrt{2} \mu \approx 125 \gev$. In \sref{ewpt} we adopt renormalization conditions to ensure that loop corrections do not change these values from their tree-level expectation. Therefore we can define the higgs Lagrangian parameters $\lambda = \frac{m_h^2}{2 v^2} \approx 0.129$ and $\mu = \frac{m_h}{\sqrt{2}} \approx 88.4 \gev$.

\subsection{Physical Parameter Space}
\label{ss.couplingplane}

\begin{figure}
\begin{center}
\hspace*{-14mm}
\begin{tabular}{cc}
\includegraphics[width=8.5cm]{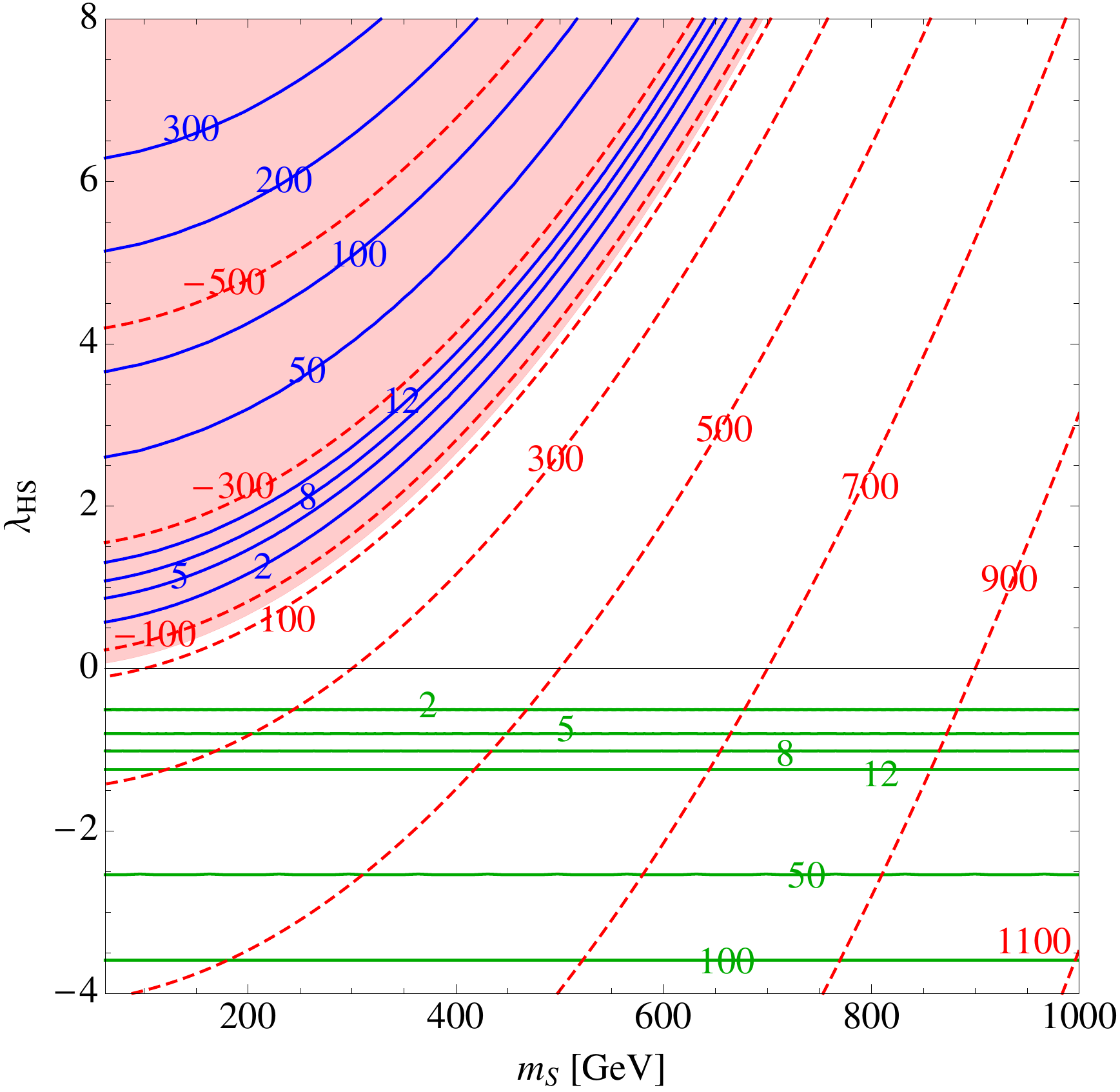}&
\includegraphics[width=8.5cm]{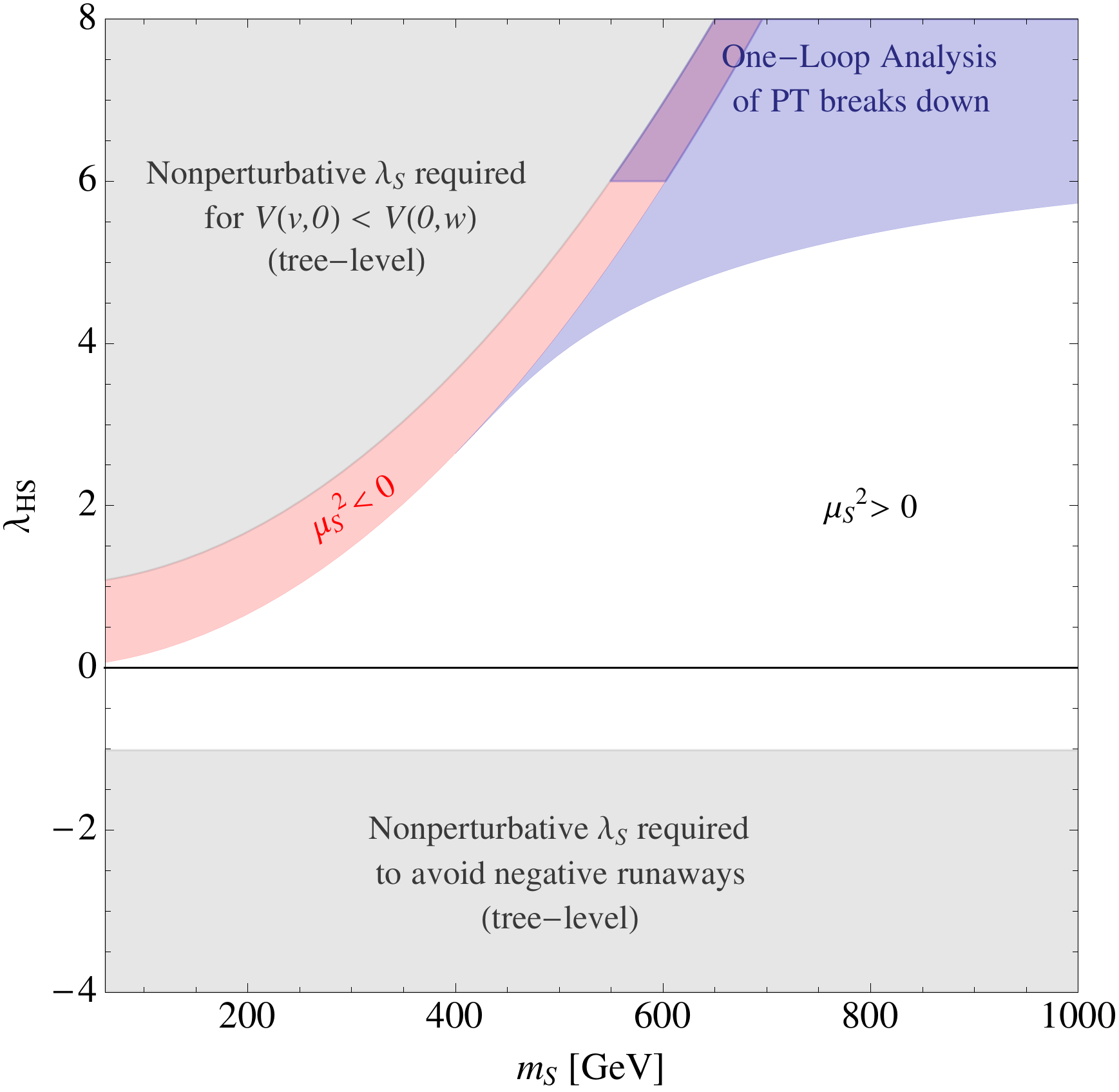}
\end{tabular}
\caption{
The parameter space of the $\mathbb{Z}_2$ symmetric SM+S extension with $m_S > m_h/2$ (our nightmare scenario). \textbf{Left:} The red shaded region indicates when $\mu^2$ is negative. The dotted red contours indicate $\mathrm{Sign}(\mu_S^2)|\mu_S|$. The blue contours show the minimum $S^4$ quartic coupling $\lambda_S$ required for the electroweak symmetry breaking (EWSB) vacuum to be the ground state of the universe, while the green contours show the minimum $\lambda_S$ to avoid negative runaways. \textbf{Right:} Gray regions indicate where theoretical control is lost due to non-perturbative $\lambda_S$. Perturbative analysis of the phase transition breaks down in the blue shaded regions, see \sref{ewpt}. The red and white regions are the possible parameter space of this nightmare scenario.}
\label{f.SMSparamspace}
\end{center}
\end{figure}

The model is determined by three new parameters, $\mu_S,\lambda_{HS}$ and $\lambda_S$.  However, in the context of our nightmare scenario, it is straightforward to show that all relevant physics can be recast into the simple two-dimensional plane of the physical singlet mass and its coupling to the higgs. 

Without excluding the possibility of a two-step phase transition where the singlet acquires a VEV at some point in cosmological history, we operate under the assumption that we live in a zero-temperature vacuum where the higgs has a VEV and the singlet does not. The mass squared of the singlet in our vacuum, required to be positive, is then 
\begin{equation}
\label{e.mSexpression}
m_S^2 = \mu_S^2 + \lambda_{HS} v^2 > 0.
\end{equation}
The other parameter which dictates the phenomenology of the singlet is its coupling to our sector through the higgs, the $hSS$ coupling. This coupling determines singlet production and annihilation cross sections and is given by $\lambda_{HS}$\footnote{When discussing the effective potential at one-loop in \sref{ewpt} we choose a scheme in which the tree-level parameter $\lambda_{HS}$ corresponds to the physical $hSS$ coupling $\mathcal{L_\mathrm{eff}} \supset -  v \lambda_{HS} h SS$.}.  The singlet self interaction, $\lambda_S$, is important when discussing regions with a possible phase transition, but does not play a direct role in the phenomenology of this model. Thus, \emph{all} the relevant features of our nightmare scenario can be shown in the $(m_S, \lambda_{HS})$ plane.

The $(m_S, \lambda_{HS})$ plane can be divided into regions where all couplings are under perturbative control or not, and further divided based on the sign of $\mu_S^2$. This division has consequences for the vacuum structure of the theory, and hence the qualitative mechanisms at play to produce strong phase transitions.  If all the quartics are positive, then for positive $\mu_S^2$ the only minimum is the EWSB vacuum at $(h,S) = (v,0)$. When $\lambda_{HS} > m_S^2/v^2$, $\mu_S^2$ is negative. This region is shaded red in \fref{SMSparamspace} (left). In this case, there are two local minima: the EWSB vacuum and a ``singlet-VEV vacuum" at $(h,S)=(0,w)$.  A surviving $\mathbb{Z}_2$ symmetry prevents higgs-singlet mixing in both vacua. 

For the scenario with negative $\mu_S^2$, we can ensure that our universe ends up in the correct EWSB vacuum by requiring that the potential $V_0(h,s)$ satisfies $V_0(0,w) > V_0(v, 0)$. It is clear that this requires a minimum value of $\lambda_S$ which depends on the choice of $m_S$ and $\lambda_{HS}$:
\begin{equation}
\label{e.lambdaSmin}
\lambda_S^\mathrm{min} = \lambda \frac{\mu_S^4}{\mu^4} = \frac{2(m_S^2 - v^2 \lambda_{HS})^2}{m_h^2 v^2}
\end{equation}
The blue contours in \fref{SMSparamspace} (left) show this minimum $\lambda_S$ at tree-level, which rapidly becomes non-perturbative as we move deeper into the shaded red region. 
Requiring $\lambda_S < 8$ excludes the gray region in the top corner of  \fref{SMSparamspace} (right) from being part of the viable parameter space.  In the remaining red strip, it is possible to choose a quartic coupling $\lambda_S$ to ensure the universe eventually ends up in the EWSB vacuum. 

There are additional constraints on $\lambda_S$ that come from avoiding runaway directions in the potential at large field values. Avoiding a negative runaway\footnote{In the presence of negative runaways the tree level potential has local minima at $h, S \neq 0$. However, by the positivity assumption of \eref{mSexpression}, these local minima are always at higher potential than the electroweak breaking minimum.} at tree-level requires 
\begin{equation}
\lambda > 0 \ \ , \ \ \ 
\lambda_S > 0  \  \ , \ \ \ 
\lambda_{HS} > - \sqrt{\lambda \lambda_S}.
\end{equation}
For a given negative $\lambda_{HS}$, which in our scenario implies $\mu_S^2 > 0$, this requirement leads to a minimum value of $\lambda_S$:
\begin{equation}
\label{e.lambdaSmin2}
\lambda_S^\mathrm{min'}  = \frac{\lambda^2_{HS}}{\lambda} = \lambda_{HS}^2 \frac{2 v^2}{m_h^2}
\end{equation}
which is indicated by the green contour lines in \fref{SMSparamspace} (left). Again, the required quartic coupling $\lambda_S$ becomes non-perturbative as we move to larger negative $\lambda_{HS}$. Applying the same $\lambda_S < 8$ cutoff as before excludes the lower gray shaded region in \fref{SMSparamspace} (right). This corresponds to the requirement that 
\begin{equation}
\lambda_{HS} \gtrsim - 1.0.
\end{equation}

In allowing $\lambda_S$ to be as big as 8 we are being somewhat generous -- theoretical control could break down at smaller couplings. However, the purpose of this demarcation of parameter space is to identify regions that we would need to probe, with either direct or indirect measurements, to exclude this model as a viable EWBG scenario. It is therefore sensible to charitably assess theoretical control and slightly over-estimate the size of parameter regions with a strong phase transition. This ensures that no viable EWBG scenarios are missed.  In particular, as we will show in future sections, the region that is explored with a more optimistic definition of perturbative control is always easier to probe directly or indirectly, thereby not changing the conclusions of our study.

The constraints that excise certain regions of parameter space thus far are based on tree-level considerations requiring couplings at non-perturbative values.  
There are additional perturbativity constraints from quantum effects. In \sref{ewpt}, we will demonstrate that a one-loop perturbative analysis of the phase transition breaks down for $\lambda_{HS} \gtrsim 6$ for $\mu_S^2 < 0$ and for $\lambda_{HS} \gtrsim 5$ for $\mu_S^2 > 0$. (This roughly coincides with regions where the quartic couplings develop Landau Poles below 10 - 100 TeV, as we discuss in \sref{strongcouplings}.) These regions, which are meant to be \emph{approximate indications} of where perturbative calculations become very unreliable, are shaded blue in the top right corner of \fref{SMSparamspace} (right).

The viable parameter space of the nightmare scenario is therefore the red and white regions in \fref{SMSparamspace} (right).  As we will see, these two regions behave very differently with regards to EWBG as well as their signals for direct and indirect measurements. The phenomenology of regions with large couplings is further discussed in \sref{strongcouplings}.

\section{Electroweak Phase Transition}\label{s.ewpt}

In this section, we will discuss the different types of phase transitions that occur in the nightmare scenario and lay out the physical parameter space in which a strong electroweak phase transition could occur. 

Successful EWBG requires a phase transition stronger than that found in the SM. This can be achieved with a variety of different mechanisms, such as thermally-driven scenarios, tree-level modifications to the scalar potential from renormalizable or non-renormalizable operators, and zero-temperature loop effects (see e.g.~\cite{Chung:2012vg}). In principle, a given model can realize several different mechanisms in different regions of its parameter space. In particular, we will demonstrate that the singlet model can have thermal, tree-level, and loop-level induced first-order EW phase transitions.  This observation is not novel, and the different mechanisms have been demonstrated individually in the literature \cite{Barger:2007im, Noble:2007kk, Espinosa:2011ax, Cline:2012hg, Cline:2013gha, Alanne:2014bra, Espinosa:2007qk}.  However, rather than simply doing a parameter scan for possible phase transitions, we examine the physics of each type of first-order phase transition, and map the effects onto the relevant phenomenological parameter space $(m_S, \lambda_{HS})$ for testing the EW phase transition. This ensures we consider every possibility for EWBG.

Before demonstrating the details of the parameter space for each type of first-order phase transition, it is useful to summarize the underlying mechanisms and how they operate in the context of this nightmare scenario. 

In \sref{scenario} we outlined how the most important order parameter separating different phases of the theory is $\mu_S^2$, the scalar mass at the origin.  The singlet potential is positive definite for $\mu_S^2 > 0$. In this case, the phase transition occurs purely along the higgs direction. However, if the singlet is sufficiently strongly coupled to the higgs, its zero-temperature loop corrections to the higgs potential can be big enough to allow SM thermal effects to trigger a strong phase transition. On the other hand, if $\mu_S^2$ is negative there can be two vacua. The universe can then undergo a two-step phase transition, first to a singlet VEV vacuum, and then to the true EWSB vacuum. This tree-level modification of the higgs potential can result in an arbitrarily strong phase transition by adjusting the potential difference between the two vacua via the choice of $\lambda_S$. If $\mu_S^2 \lesssim - (100 \gev)^2$ and the singlet self- and scalar-couplings are just right, it is also possible for a one-step transition to occur via thermal effects, akin to the MSSM light stop scenario. This is only realized for a small region of parameter space, entirely contained within the two-step phase transition region of the $(m_S, \lambda_{HS})$ plane. It is therefore clear that the two different regions of parameter space delineated in \sref{scenario} also realize strong phase transitions differently: one with relatively low $m_S$ and negative $\mu_S^2$, and one with large $m_S$ and stronger singlet-higgs coupling.

In what follows, we review the phase transition calculation in detail and define the regions of parameter space which can realize EWBG. It is important to note that while these regions are very distinct from the point of view of the phase transition, they are continuously connected in the phenomenological parameter space $(m_S, \lambda_{HS})$.

\subsection{$\mu_S^2 > 0$: One-Step Transition via Loop Effects}
\label{ss.onestep}

For $\mu_S^2 > 0$ and without negative runaways, the singlet never attains a VEV, and there are no tree-level effects to enhance the phase transition.  However, it is still possible to induce a strong electroweak phase transition via sizable one-loop zero-temperature corrections to the SM higgs potential.

\subsubsection{Effective Potential}
\label{sss.Veff}

The finite-temperature effective higgs potential \cite{
Cline:2006ts, Trodden:1998ym, Riotto:1998bt, Riotto:1999yt, Quiros:1999jp, Morrissey:2012db} is made up of four components:
\begin{equation}
\label{e.Veff}
V_\mathrm{eff}(h,T)=V_0(h)+V_0^{CW}(h)+V_T(h,T)+V_r(h,T).
\end{equation}
$V_0$ is the tree-level potential defined in \eref{V0}. $V_0^{CW}$ is the one-loop zero-temperature correction \cite{Coleman:1973jx} , $V_T$ is the one-loop finite-temperature potential, and $V_r$ are the ring-terms. See \aref{FTEP} for the full expressions.

The singlet quartic $\lambda_S$ does not contribute to any mass term when $S = 0$. 
In fact, its sole appearance is in the zero-momentum polarization tensor $\Pi_S(0)$. 
This only affects $V_r(h,T)$ and has only a very minor effect on the one-step phase transition, as we will see below. 
Therefore, at one loop, the strength of the phase transition for $\mu_S^2>0$ is almost entirely determined by the two parameters $(m_S, \lambda_{HS})$.

\subsubsection{Electroweak Phase Transition via Loop Effects}
\label{sss.ewptvialoop}

We compute the total $V_\mathrm{eff}(h, T)$ in \eref{Veff} for different choices of $(m_S, \lambda_{HS})$ in the white region of \fref{SMSparamspace} (right).\footnote{The imaginary part of $V_\mathrm{eff}$ is a spurious artifact of the perturbative expansion and is ignored \cite{Martin:2014bca}.} We set $\lambda_S = 0$; increasing it to $\lambda_S \sim \mathcal{O}(1)$ slightly weakens the phase transition, so setting the self-coupling to zero shows the largest possible region where EWBG can occur. Varying the temperature, we find $T = T_c$ where the two local minima $h = 0 $ and $h = v_c$ are degenerate, and check the ratio $v_c/T_c$ to see whether EWBG is possible according to \eref{bnpc}.\footnote{The ratio $v_c/T_c$ is not gauge invariant, and obtaining an explicitly gauge-invariant baryon-number preservation criterion requires special care to obtain a fully consistent perturbative expansion for the quantities $v_c$ and $T_c$ separately, but the numerical impact of using the fully gauge invariant criterion is much smaller than the effect of 2-loop corrections \cite{Patel:2011th}.} 

The result is shown in \fref{vcoverTcplot}. Orange contours show the value of $v_c/T_c$, with orange shading indicating the region $v_c/T_c > 0.6$ where EWBG could proceed with an efficient baryon number generation mechanism. However, the exact choice of the minimum $v_c/T_c$  does not qualitatively affect the definition of the EWBG-compatible region. Strong couplings $\lambda_{HS} \gtrsim 2$ are needed. The critical temperature decreases with increasing coupling and is in the range of $T_c \sim 130 - 160 \gev$.

To understand this mechanism for generating a strong phase transition, we repeat the above calculation with various contributions to the total effective potential \eref{Veff} switched off. We find that using
\begin{eqnarray*}
V_\mathrm{eff} &=& V_0 \ \ \ \mbox{(tree-level potential)} \\
&& +  \ \mbox{(only \emph{singlet} contributions to $V_0^\mathrm{CW}$)}\\
&& +  \ \mbox{(only \emph{SM} thermal contributions)}
\end{eqnarray*}
gives a very similar result, with the $\lambda_{HS}$ necessary for a strong phase transition underestimated by about 10\%. This implies that sizable zero-temperature one-loop higgs potential contributions from the singlet reduce the potential difference between the EWSB vacuum and the origin, which then makes it easier for SM thermal contributions to generate an energy barrier between the two degenerate local minima at some $T = T_c$. This is illustrated in \fref{Vcomparisonplot}.

For very strong coupling, the one-loop effects create an energy barrier even at zero temperature. This is the case above the dashed green line in \fref{vcoverTcplot}. However, as we discuss in the next subsection, our one-loop analysis may not be valid for such high coupling.

\begin{figure}
\begin{center}
\begin{tabular}{c}
\includegraphics[width=8.5cm]{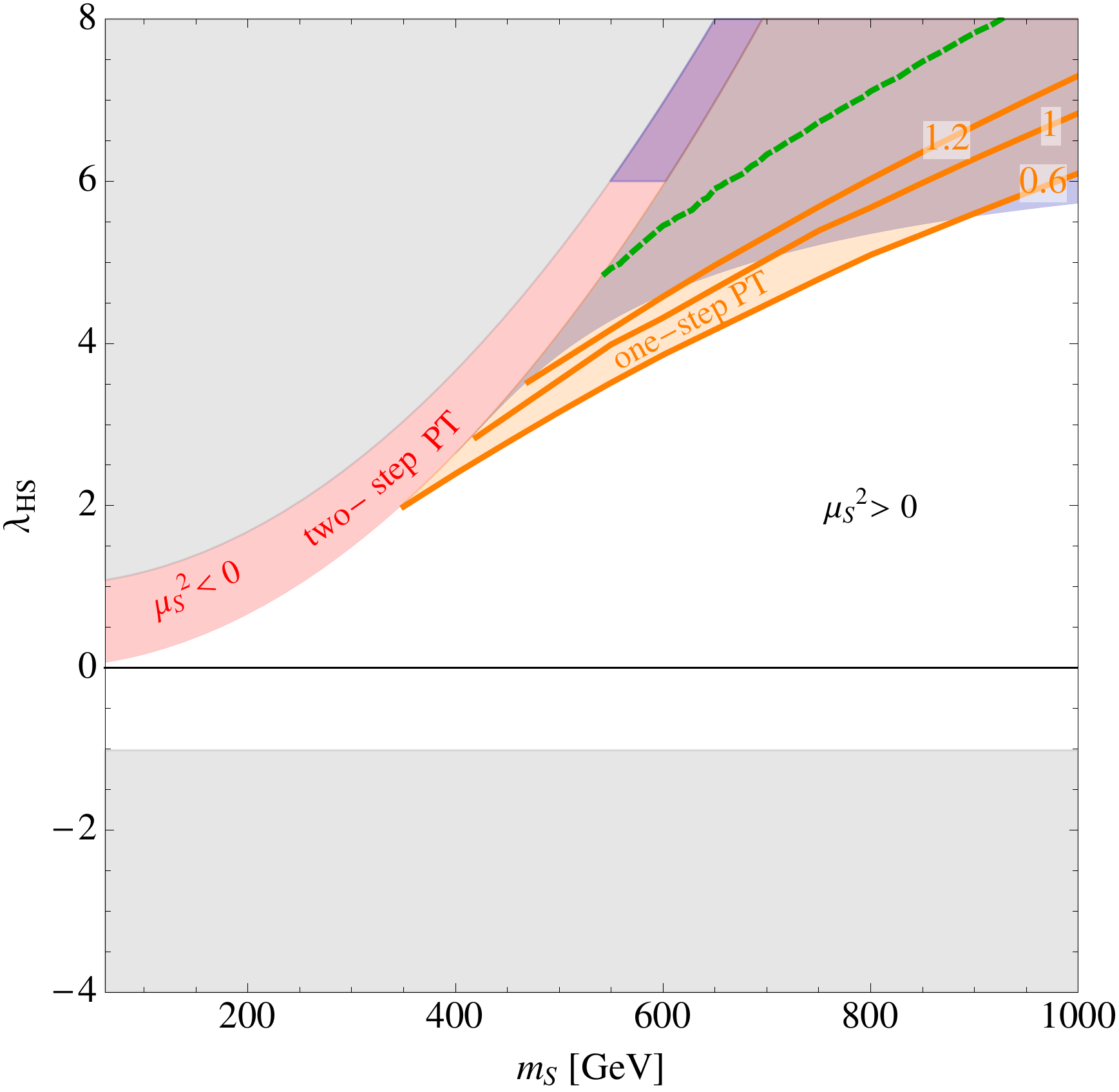}
\end{tabular}
\caption{
Regions in the $(m_S, \lambda_{HS})$ plane with viable EWBG. Red shaded region: for $\mu_S^2 < 0$ it is possible to choose $\lambda_S$ such that EWBG proceeds via a tree-induced strong two-step electroweak phase transition (PT). Orange contours: value of $v_c/T_c$ for $\mu_S^2 > 0$. The orange shaded region indicates $v_c/T_c > 0.6$, where EWBG occurs via a loop-induced strong one-step PT.  Above the green dashed line, singlet loop corrections generate a barrier between $h = 0$ and $h = v$ even at $T = 0$, but results in the dark shaded region might not be reliable, see \sssref{perturbativereliability}.
 } 
 \vspace{8mm}
\label{f.vcoverTcplot}
\end{center}
\end{figure}

\begin{figure}
\begin{center}
\hspace*{-13mm}
\includegraphics[width=0.7\textwidth]{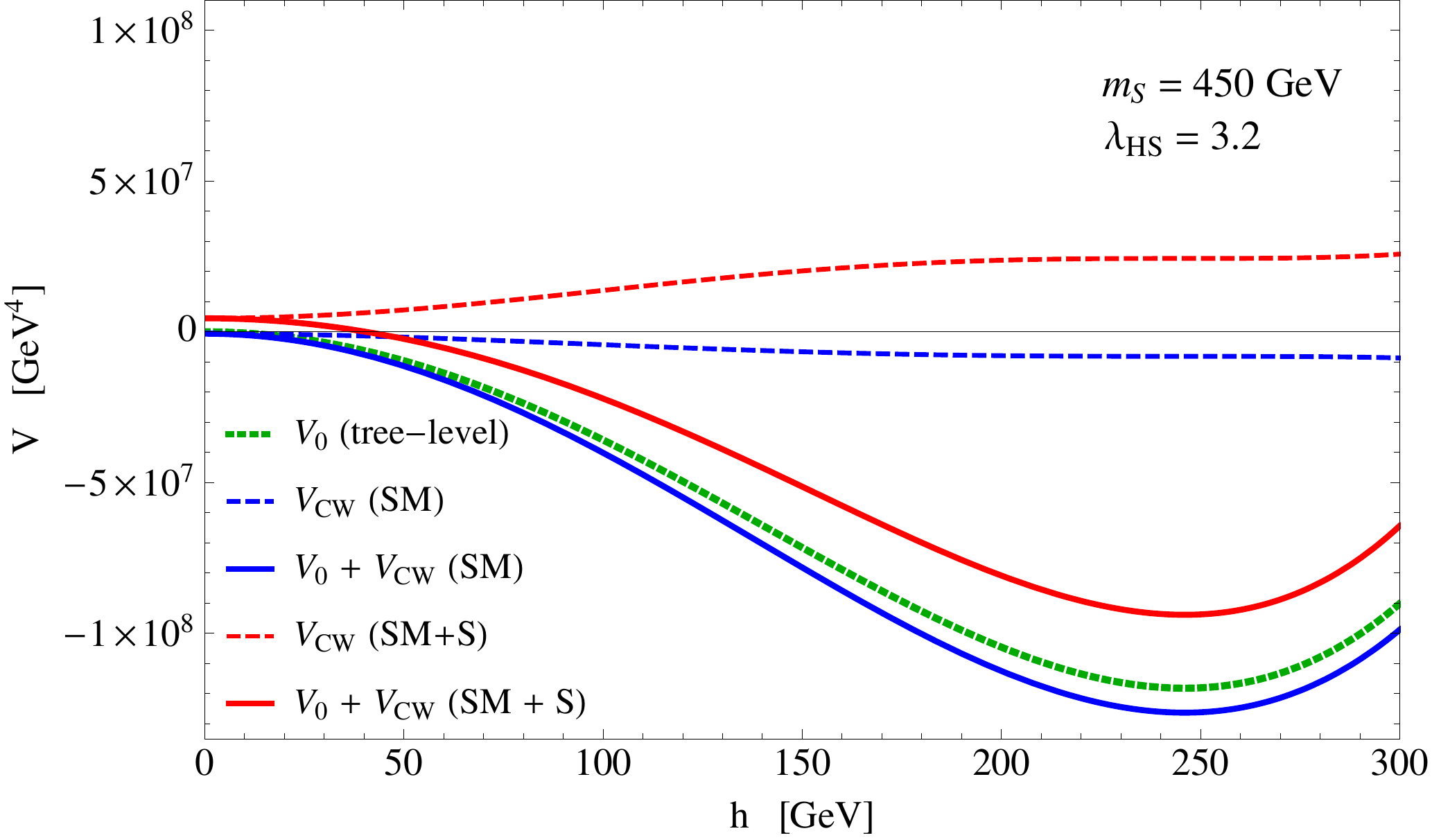}
\end{center}
\caption{
Comparison of the zero-temperature potential contributions in the SM vs. the SM + singlet with $(m_S, \lambda_{HS}) = (450 \gev, 3.2)$ which has a strong first-order PT with $v_c/T_c > 1$. The one-loop contribution of the singlet reduces the potential difference between the origin and the EWSB vacuum.
}
 \vspace{8mm}
\label{f.Vcomparisonplot}
\end{figure}

\subsubsection{Reliability of Perturbative Analysis}
\label{sss.perturbativereliability}

We have found that a strong one-step electroweak phase transition requires rather large quartic singlet-higgs couplings $\lambda_{HS} \gtrsim 2$. It is prudent to examine the validity of the perturbative expansion to understand the trustworthiness of this result. In this discussion we only consider zero-temperature loop-effects, since those are the singlet contributions responsible for a strong phase transition. 

The quartic term $h^4$ in the one-loop improved zero-temperature effective potential can be written as
\begin{equation}
V_0 + V^\mathrm{CW}_0 \supset \frac{1}{4} \left[ \lambda + \Delta \lambda  (h) \right] h^4.
\end{equation}
The one-loop singlet contribution to $\Delta \lambda$ is
\begin{equation}
\label{e.deltalambda}
\Delta \lambda(h) = \frac{\lambda_{HS}^2}{16 \pi^2} \left(  \log\left[ 1 + \frac{(h^2 - v^2)\lambda_{HS}}{m_S^2} \right] - \frac{3}{2}\right)
\end{equation}

\begin{figure}
\begin{center}
\begin{tabular}{c}
\includegraphics[width=8.5cm]{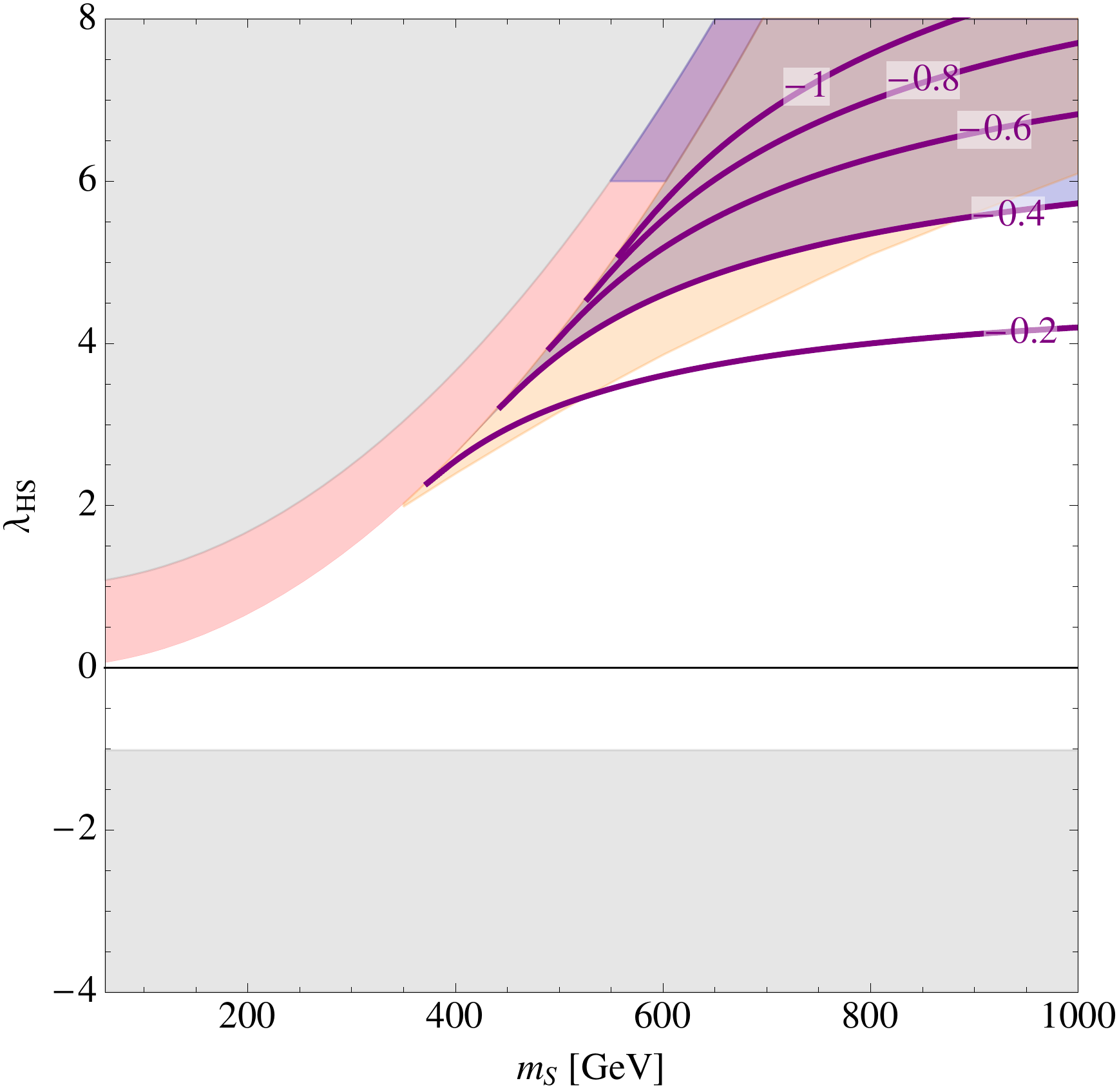}
\end{tabular}
\caption{
Same as \fref{vcoverTcplot}, but with contours of $\Delta \lambda(h=0)$ \eref{deltalambda} shown in purple. Dark shading above $|\Delta \lambda| = 0.4$ indicates approximately where the results of our analysis in \sssref{ewptvialoop} are not trustworthy due to loss of perturbativity.
}
\label{f.perturbativevalidityplot}
\end{center}
\end{figure}

Two-loop corrections scale as $\sim (\Delta \lambda)^2$, and thus the validity of the perturbative analysis requires $\Delta \lambda$ not to be too large.\footnote{Similar, though less stringent, constraints on perturbativity are obtained by considering the correction to other terms in the potential.}

\fref{perturbativevalidityplot} shows contours of $\Delta \lambda(h=0)$ in the $(m_S, \lambda_{HS})$ parameter space. The correction is evaluated at the origin to maximize its size and give a somewhat pessimistic estimate of where our perturbative analysis is trustworthy. 

For large couplings $\lambda_{HS}$, $\Delta \lambda(0)$ rapidly approaches unity. While it is difficult to quantitatively define an exact region where the analysis becomes unreliable, clearly the results for $\lambda_{HS} \gtrsim 5$ should be taken with a grain of salt. We choose the $|\Delta \lambda| = 0.4$ contour in \fref{perturbativevalidityplot} as the approximate boundary of our regime of perturbative validity, and indicate larger values with blue shading in all plots (see also \fref{SMSparamspace}).  We conclude that for $\lambda_{HS} \lesssim 4-5$, zero temperature loop effects can induce a strong electroweak phase transition and the calculation can be trusted.

We finish this discussion with a parenthetical remark. One could  think of quantifying a degree of ``fine-tuning'' by the size of $\Delta \lambda$. Given that the zero-temperature quartic of the higgs potential needs to be $\co{0.1}$, one might require $\Delta \lambda$ to ``naturally'' be of similar size, otherwise the new sector at one-loop dominates the tree-level higgs potential.   
Of course, given the contours shown in \fref{perturbativevalidityplot}, this more restrictive naturalness requirement only serves to greatly reduce the available parameter space for a strong phase transition, and as such makes testing EWBG even easier without introducing a fixed measure for ruling it out.

\subsection{ $\mu_S^2 < 0$: Two-Step Transition via Tree-Effects}
\label{ss.twostep}

It has long been understood that singlet extensions of the SM can lead to tree-level modifications of the higgs potential, creating a barrier between local minima  $h = 0$ and $h = v$. This barrier makes the electroweak phase transition strongly first-order without requiring particular quantum or thermal effects, as is the case for our model when $\mu_S^2 < 0$. 

We first explain how this occurs using simple tree-level arguments before confirming this picture with a full one-loop analysis. 

\subsubsection{Tree-level argument}

\label{sss.twostep_tree}

In the red region of \fref{SMSparamspace} (right), $\mu_S^2 < 0$, and we can choose a $\lambda_S > 0$ such that there are two local minima, one at $h = 0, S = w$ and a deeper one at $h = v, S = 0$ (at zero temperature). When the universe is very hot $T \gg 100 \gev$, thermal contributions stabilize both fields at the origin. Since the singlet couples to fewer degrees of freedom, its thermal mass is lower than that of the SM higgs. Therefore, as the universe cools, the singlet gets destabilized before the higgs (see e.g.~\cite{Cline:2012hg}). The electroweak phase transition then starts in the singlet-VEV minimum and ends in the EWSB minimum. 

As outlined in \ssref{couplingplane}, we can always choose a $\lambda_S$ to make the two local minima degenerate. This corresponds to zero critical temperature, i.e. the universe never transitions from the singlet-VEV minimum to the EWSB minimum. For any given point $(m_S, \lambda_{HS})$ in the red region of \fref{SMSparamspace} (right) one can then imagine taking $\lambda_S$ a little bit larger than $\lambda_S^\mathrm{min}$ in \eref{lambdaSmin}. This gives an arbitrarily low $T_c$, and hence an arbitrarily large ratio $v_c/T_c$, easily satisfying the baryon number preservation criterion \eref{bnpc}, while ensuring the singlet-VEV vacuum is short-lived.\footnote{
The tunneling rate from the singlet-VEV minimum to the EWSB minimum is $\Gamma \sim e^{-S_E}$, where $S_E$ is the finite-temperature bounce action \cite{Coleman:1977py}. For $S_E \sim 100$, the false vacuum decays quickly in the early universe \cite{Delaunay:2007wb}. We computed the \emph{zero temperature} bounce action $B$ in the triangle potential barrier approximation \cite{Duncan:1992ai} and found that $B < 100$ for some range of  $\lambda_S < 8$ in most of the red shaded region of \fref{SMSparamspace}. Thermal fluctuations greatly enhance the tunneling rate, $S_E < B$. Therefore, the transition between the two minima can be sufficiently fast to ensure a viable thermal history for the universe.} 

The above discussion may be modified slightly by loop and thermal effects. By and large, however, in (or close to) the red shaded region of \fref{SMSparamspace} (right) and \fref{vcoverTcplot}, EWBG is possible via a strong two-step phase transition, which is induced by tree-level modifications to the higgs potential.

\subsubsection{Full Analysis}
\label{sss.twostep_full}

We confirm the validity of the above argument with an explicit calculation. The two-dimensional effective potential $V_\mathrm{eff}(h,S,T)$ is obtained from \eref{Vcw} by including the singlet-dependence of the singlet and higgs masses:
\begin{equation}
m_h^2 = -\mu^2 + 3 h^2 \lambda + \lambda_{HS} S^2 \ \ , \ \ \ \ 
m_S^2 = \mu_S^2 + 3 \lambda_S S^2 + \lambda_{HS} h^2.
\end{equation}

The first step is finding the minimum value of $\lambda_S = \lambda_S^\mathrm{min}(m_S, \lambda_{HS})$ required to satisfy the condition 
\begin{equation}
V_\mathrm{eff}(0,w,T=0) > V_\mathrm{eff}(v,0,T=0).
\end{equation}
Requiring $\lambda_S^\mathrm{min}(m_S, \lambda_{HS}) < 8$ at tree-level was used in  \ssref{couplingplane} to define the viable $\mu_S^2 < 0$ region of parameter space, shaded red in Figs.~\ref{f.SMSparamspace}~(right) and \ref{f.vcoverTcplot}. We find that the definition of this region does not change significantly when including loop corrections, except for the fact that $\lambda_{S}^\mathrm{min} > 8$ for all $\mu_S^2 < 0$ when $\lambda_{HS} \gtrsim 6$. Therefore we regard any calculation in the $\mu_S^2 < 0$ region with larger $\lambda_{HS}$ coupling as unreliable, indicated with the blue shading above $\lambda_{HS}  = 6$ in \fref{SMSparamspace} (right) and all following figures. 

A given choice of $(m_S, \lambda_{HS})$ and $\lambda_S > \lambda_S^\mathrm{min}(m_S,\lambda_{HS})$ completely defines the  temperature-dependent effective potential, and it is straightforward to analyze the two-step phase transition. We will focus on values of $\lambda_S = \lambda_S^\mathrm{min} + \mathcal{O}(0.1)$. 

At very high temperature both fields $(h,S)$ are stabilized at the origin\footnote{This is not the case for large $\lambda_S$ or $\lambda_{HS}$, since the singlet thermal potential develops a high-temperature instability if $\frac{1}{3}\lambda_{HS} + \frac{1}{3}\lambda_S > \frac{\pi^2}{9} \approx 1.1$. At high temperature, the singlet then has nonzero VEV. This does not affect our argument for a strong phase transition, since it essentially corresponds to  $T_{c1} \gg T_{c2}$. It also does not affect the one-step phase transition for $\mu_S^2 > 0$.}.  As the universe cools, the singlet transitions to a nonzero VEV first, in a second-order phase transition at temperature $T_{c1} \sim $ few $100 \gev$. The EWSB minimum eventually drops below the singlet-VEV minimum at temperature $T_{c2} < T_{c1}$. Since the tree-level barrier between the two minima survives at $T_{c2}$, the universe undergoes a first-order phase transition to $(v_{c2},0)$, where $v_{c2} \equiv v(T_{c2})$. $T_{c2}$ varies in the 2-step phase transition region, increasing as we take $\mu_S^2 \rightarrow 0$. 
\begin{itemize}
\item For $\lambda_{S} = \lambda_S^\mathrm{min}$, $T_{c2} < 45 \gev$ in the entire two-step region. The phase transition is very strong, $v_{c2}/T_{c2} \sim 4$ near $\mu_S^2 \to 0$ (outer boundary of 2-step phase transition region) and $\gtrsim 8$ as $\lambda_{S}^\mathrm{max}$ approaches its maximum allowed value of 8 (inner boundary of 2-step phase transition region). 
\item For $\lambda_S = \lambda_S^\mathrm{min} + 0.1$, $T_{c2} \sim 30 - 100 \gev$ in the entire 2-step phase transition region, with $v_{c2}/T_{c2} > 2$.
\end{itemize}
Clearly a relatively small increase in $\lambda_S$ compared to its minimum value guarantees that the phase transition takes place at a cosmologically safe temperature, and the singlet-VEV vacuum is short-lived. 

This calculation demonstrates the validity of the tree-level arguments in \sssref{twostep_tree}. A strong two-step phase transition can be achieved in the entire viable $\mu_S^2 < 0$ region, shaded red in all our plots. The loop-level analysis reveals perturbativity is lost for $\lambda_{HS} \gtrsim 6$, which is shaded blue in all our plots.

\subsection{ $\mu_S^2 < 0$: One-Step Transition via Thermal Effects}
\label{ss.thermal}

It was found previously \cite{Katz:2014bha} that an unmixed singlet extension of the SM with a \emph{complex} scalar could, for $\mu_S^2 < 0$ and sizable coupling to the higgs, induce a strong one-step phase transition for some choice of self-coupling which stabilizes the singlet at the origin when $T = T_c$.

We find that this mechanism can also be realized in our model, which only has a single \emph{real} scalar. In parts of our $\mu_S^2 < 0$  two-step phase transition region, for some choices of $\lambda_S > \lambda_S^\mathrm{min}$ and $|\mu_S| \lesssim 100 \gev$, the singlet bare mass cancels its thermal mass and generates a negative cubic term in the finite-temperature higgs potential, while also stabilizing the singlet at the origin for $T \geq T_c$. This replicates the well-known mechanism for a strong phase transition realized, for example, by the Light Stop Scenario in the MSSM~\cite{
Huet:1995sh, Carena:1996wj, Laine:1998qk, Laine:2000xu, Espinosa:1996qw, Carena:1997ki, Huber:2001xf, Cline:2000kb, Carena:2002ss, Lee:2004we, Cirigliano:2009yd, Carena:2008rt, Carena:2008vj, Quiros:1999tx, Delepine:1996vn}.

However, since the thermally driven phase transition only occurs for some finely tuned $\lambda_S$ in a very small part of the two-step phase transition region, we do not miss any EWBG-viable regions by only discussing the $\mu_S^2 > 0$ one-step and $\mu_S^2 < 0$ two-step phase transition regions in the phenomenological analysis of the following sections.

\subsection{Summary}

\fref{vcoverTcplot} shows the two regions in the nightmare scenario's parameter space where EWBG is possible. For $\mu_S^2 < 0$, a judicious choice of $\lambda_S$ can always generate a strong two-step phase transition via tree effects (and sometimes thermal effects) in the red-shaded region. For $\mu_S^2 > 0$, zero-temperature loop effects from the singlet raise the EWSB minimum, which allows SM thermal contributions to generate a sizable energy barrier. This makes EWBG possible in the orange-shaded region. For $\lambda_{SH} \lesssim 5$ $(6)$ in the one (two) step phase transition region our analysis is perturbatively reliable, with untrustworthy regions indicated by the blue shading in all our figures.

This establishes the regions of the $(m_S, \lambda_{HS})$ plane where EWBG could occur. We now move on to discuss ways of directly and indirectly detecting signatures of the strong phase transition.

\section{Direct Signatures of the Phase Transition}\label{s.collider}
By construction, the only way to directly produce singlets in the nightmare scenario is through pair production via an off-shell higgs. Since the singlets are observed as missing energy, a visible object needs to be produced in association with the singlets in order to discover them. Given that the only coupling of the singlets to the visible sector is through the higgs, standard invisible higgs channels are potentially useful to examine: monojet, associated production (AP), and vector boson fusion (VBF). The main differences are the unknown invariant mass of the final state, and a much smaller cross section. Monojet searches are the most difficult given the QCD background, but dedicated investigations may yield some reach in this channel ~\cite{NateMattArun}. The cleanest channels in which to search for the singlet are AP $q\bar q\to V S S$, and VBF $qq\to  SSqq$, due to leptonic final states and distinctive kinematics of the jets, respectively.

Cross-sections for AP and VBF, shown in \fref{xsecSS}, are very small even at a 100 TeV collider. This makes direct searches very challenging. Given that the VBF channel has the largest cross section we use it as a litmus test for a putative 100 TeV direct search strategy.  In principle, combining AP, VBF and monojet searches could improve the reach somewhat~\cite{NateMattArun}, but the qualitative lessons we demonstrate below will hold.

\begin{figure}
\subfigure[~$\sqrt s = $ 14 TeV]{\includegraphics[width=0.5\textwidth]{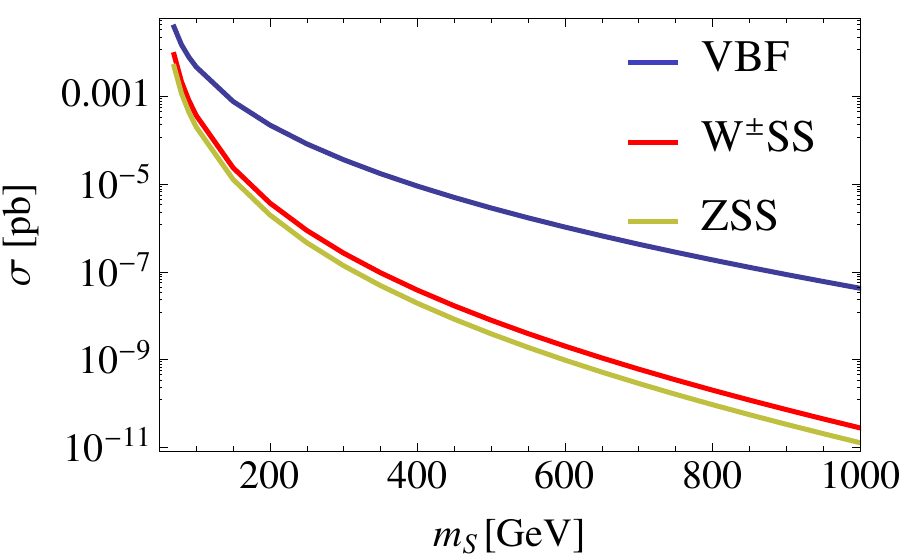}}
\subfigure[~$\sqrt s = $ 100 TeV]{\includegraphics[width=0.5\textwidth]{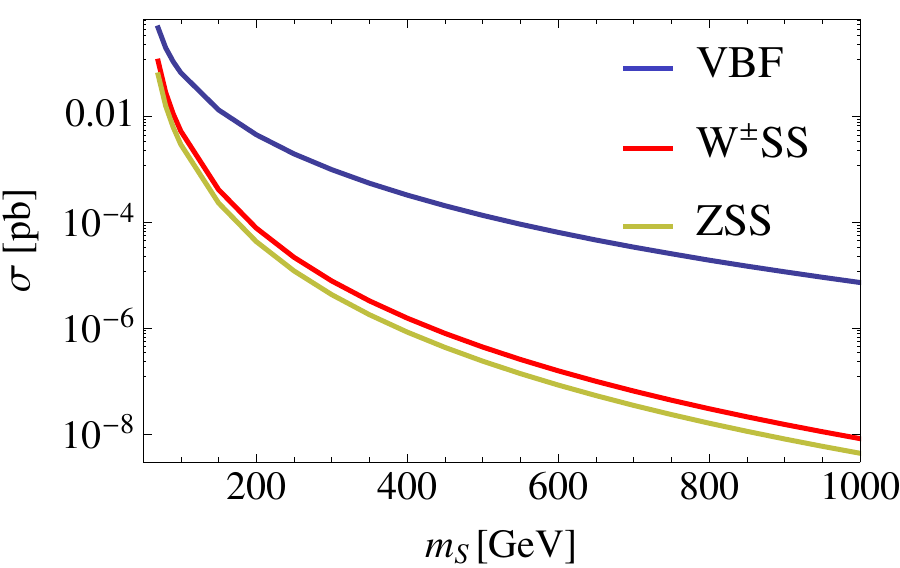}}
\caption{Production cross-sections at hadron colliders for various modes of singlet production with $\lambda_{HS}=2$. These calculations were computed at LO with MadGraph5 \cite{Alwall:2011uj}}
\vspace{12mm}
\label{f.xsecSS}
\vspace{8mm}
\end{figure}

The dominant background for VBF singlet production (with a moderate missing energy requirement) is $(Z \to \nu \nu)$ + jets. 
The VBF production cross section of $Z\to \nu\nu$ is around 1000~pb for a 100 TeV $pp$ collider. This is already much larger than the $< 10^{-2}$ pb for VBF production of $h\to SS$, and does not include non-VBF $Zjj$.
Despite these discouraging numbers, we will show it is still possible to have sensitivity to the parameter space relevant for EWBG at a 100 TeV collider.

To see this, we consider a simple VBF analysis with the following criteria:
\begin{itemize}
\item exactly two jets with $p^T_{j_{1,2}} > 40$ GeV, $|\eta_{j_{1,2}}| < 5$
\item  $\met>$ 150 GeV,
\item $\Delta\eta_{jj}=|\eta_{j_1}-\eta_{j_2}|>3.5$ and $|\eta_{j_{1,2}}| > 1.8$,
\item $M_{jj}>800$ GeV.
\item reject events with leptons satisfying $|\eta| < 2.5$ and $p_T > 15 \gev$. 
\end{itemize}
We consider $(Z\to \nu \nu) + jj$ background from both Drell-Yan and VBF production. 
We use \texttt{MadGraph5 v1.5.12}~\cite{Alwall:2011uj} evaluated with the \texttt{CTEQ6l}~\cite{Pumplin:2002vw,Nadolsky:2008zw} parton distribution functions and \texttt{Pythia8}~\cite{Sjostrand:2006za,Sjostrand:2007gs} showering \& hadronization to generate the signal events. For detector simulation, we use \texttt{Delphes 3.1.2}~\cite{deFavereau:2013fsa} with the same detector card as the 100 TeV Snowmass Studies \cite{Anderson:2013kxz,Avetisyan:2013onh,Avetisyan:2013dta}.  For the background, we used pre-computed \texttt{Bj-4p} and \texttt{Bjj-vbf} event samples without pile-up from the Snowmass database~\cite{snowmassbkgd}. Pile-up was neglected. \fref{SB100TeV} shows the resulting $S/\sqrt{B}$ contours in the $(m_S, \lambda_{HS})$ plane for a 100 TeV $pp$ collider with $3 \iab$ and $30 \iab$ of data. 

\begin{figure}[htb]
\centering
\hspace*{-14mm}
\begin{tabular}{cc}
\includegraphics[width=8.5cm]{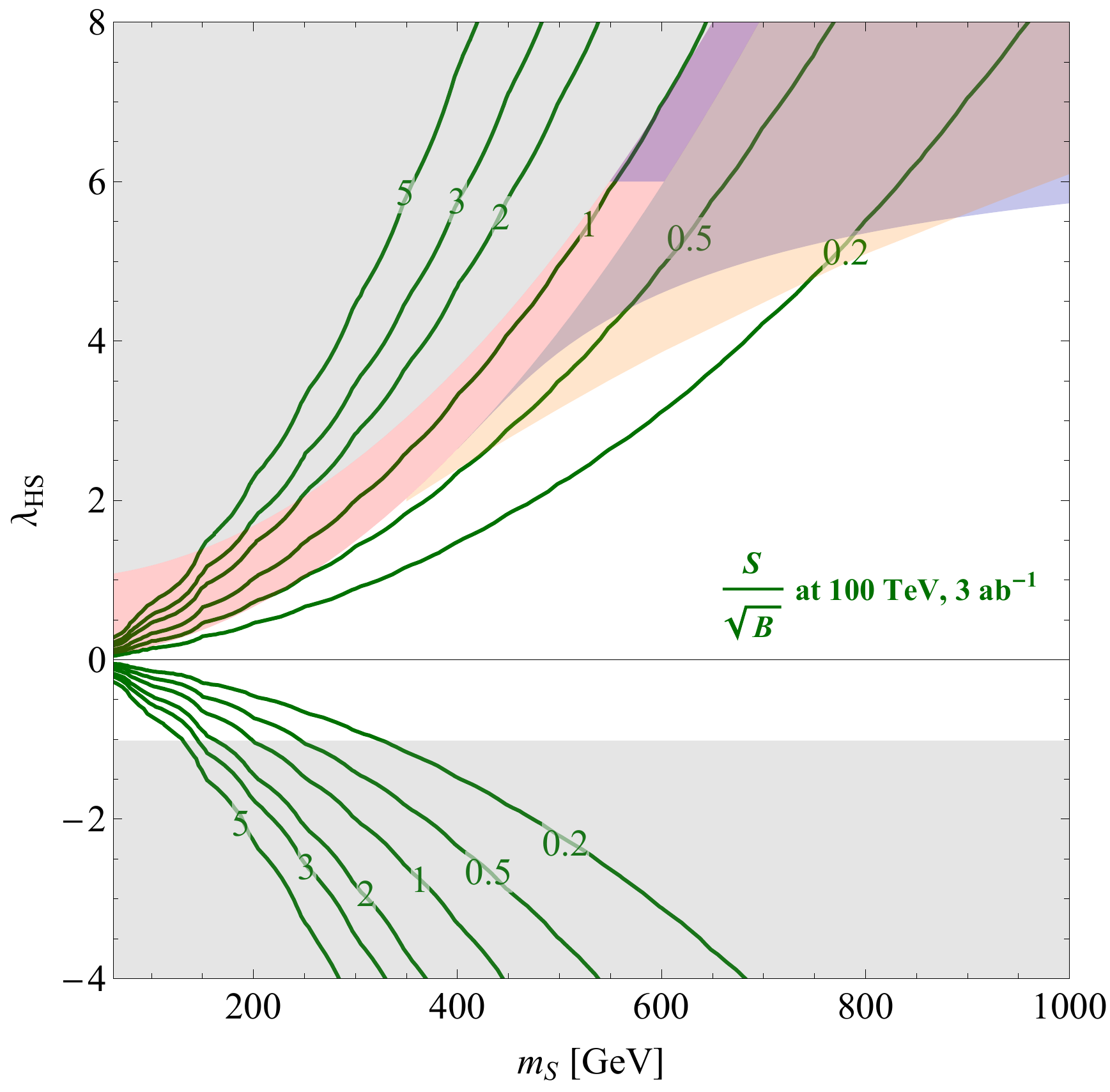}
\includegraphics[width=8.5cm]{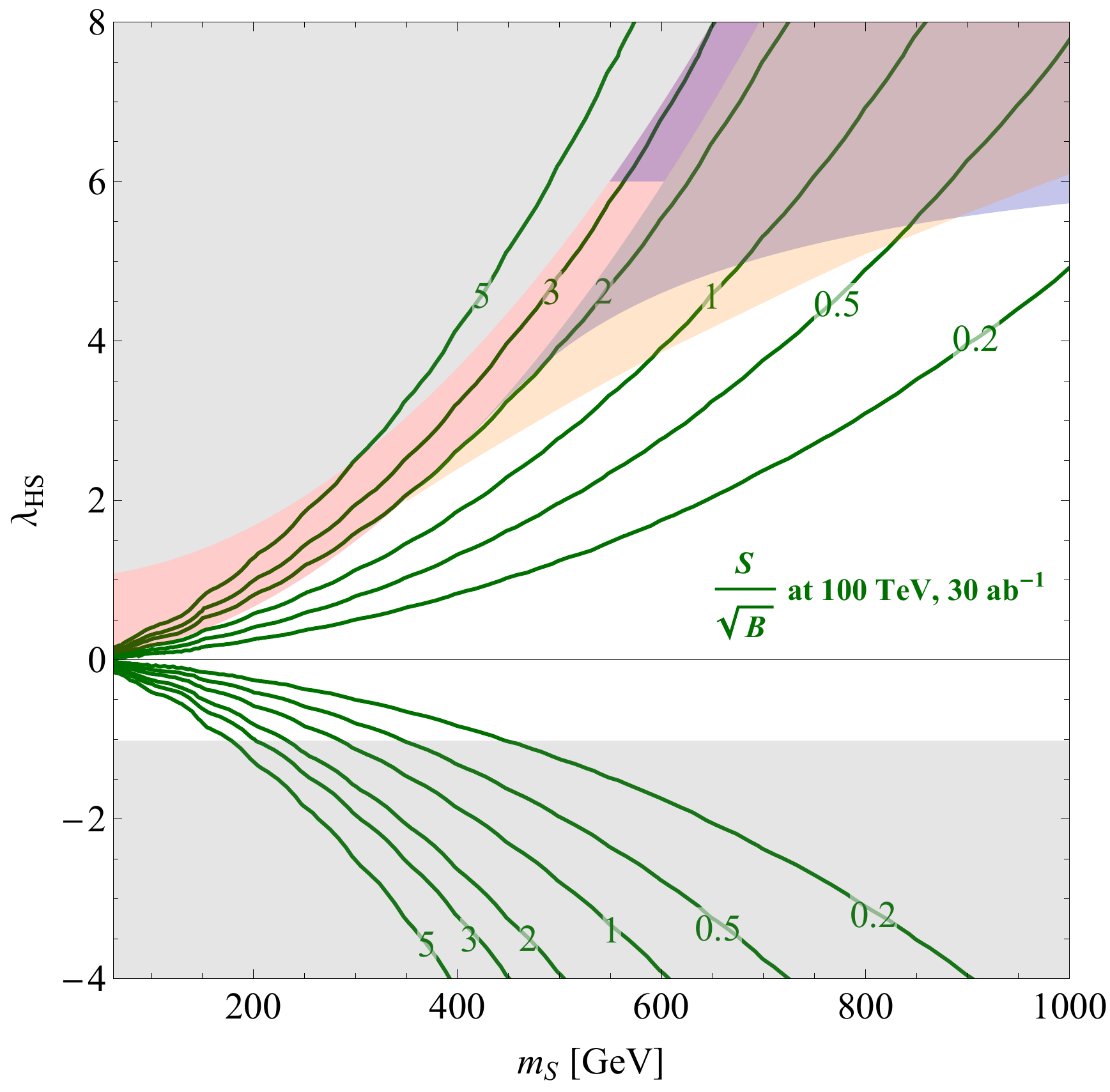}
\end{tabular}
\caption{
Green contours show $S/\sqrt{B}$ for VBF production of the $SSqq$ signal vs main background, $(Z\to\nu\bar \nu) + jj$, for a 100 TeV $pp$ collider with $3 \iab$ (left) and $30 \iab$ (right) of data. VBF selection criteria and a $\met>150$ GeV requirement were used to cut down on QCD background. Shading identical to Figs.~\ref{f.vcoverTcplot} and \ref{f.perturbativevalidityplot}.
}
\vspace{8mm}
\label{f.SB100TeV}
\end{figure}


Our naive estimate suggests $S/\sqrt{B}$ is order unity in the entire two-step phase transition region. The actual sensitivity will depend on the detector capabilities and total luminosity of the potential future 100 TeV collider program, but probing the entire two-step region via singlet VBF production at the $95\%$ confidence level may be possible, especially with $30 \iab$. More sophisticated search and background reduction techniques may improve on these estimates.

This search will be challenging in practice due to its sensitivity to systematic errors. However, there are potential data-driven methods for addressing this. For example, the $(Z\to\nu\nu)jj$ background is kinematically identical to the $(Z\to\ell\ell)jj$ background under the replacement of $p^T_{\ell\ell} \ \to\  \met$. This suggests a very statistically precise background template could be derived from data, greatly reducing systematics compared to a naive estimate.

Most of the parameter space for the strong one-step phase transition seems entirely out of reach by direct detection. However, as we see below, indirect measurements can be sensitive to the rest of the relevant parameter space.

\section{Indirect Signatures of the Phase Transition}\label{s.indirect}

As we saw in Sec.~\ref{s.collider}, direct searches at a 100 TeV collider can probe the two-step but not the one-step phase transition region. However, indirect searches have very complementary reach and are a promising avenue for detection. Past works using EFT formulations ~\cite{Zhang:1993vh, Grojean:2004xa, Delaunay:2007wb} and complex singlets ~\cite{Katz:2014bha} have shown a strong connection between a strong first-order phase transition and shifts in the triple higgs coupling or the $Zh$ cross-section. However, these results are not directly applicable to our model. The EFT formulation describes a different type of phase transition than what we consider and maps poorly onto our theory.  On the other hand, \cite{Katz:2014bha}  studied only thermally driven transitions, and only in models with more than one real scalar degree of freedom with large couplings. 

This lends credence to our label of a ``nightmare scenario'' for the model we study, since a strong phase transition can occur with much weaker indirect collider signatures than in the above two examples. However, it will still be testable with certain future colliders.

\subsection{Triple-higgs Coupling}
\label{ss.triplehiggs}

The triple-higgs coupling in our EWSB vacuum $\langle h \rangle = v, \langle S \rangle = 0$ is related to the third derivative of the zero-temperature effective potential 
\begin{equation}\left.
\lambda_3\equiv \frac{1}{6}\frac{d^3\left(V_0(h)+V_0^{CW}(h)\right)}{dh^3}\right|_{h=v} = \frac{m_h^2}{2 v} + \frac{\lambda_{HS}^3 v^3}{24 \pi^2 m_S^2} + \ldots
\end{equation}
The first and second term above is the SM tree-level and singlet loop-level contribution. Other subdominant SM loop contributions are not shown.  \fref{lambda3} shows $\lambda_3/\lambda_3^\mathrm{SM}$ in the $(m_S, \lambda_{HS})$ plane. For illustrative purposes, the contours are also shown in the areas where $\lambda_S$ is non-perturbative. 

As pointed out by \cite{Noble:2007kk}, a strong one-step phase transition via the effects of a real singlet is correlated with a large correction to $\lambda_3$. \fref{lambda3} shows that requiring $v_c/T_c > 0.6$ (1.0) implies $\lambda_3/\lambda_3^\mathrm{SM} > 1.2$ $(1.3)$.  Such a sizable deviation makes it possible to exclude this type of strong phase transition.

\begin{figure}
\centering
\includegraphics[width=8.5cm]{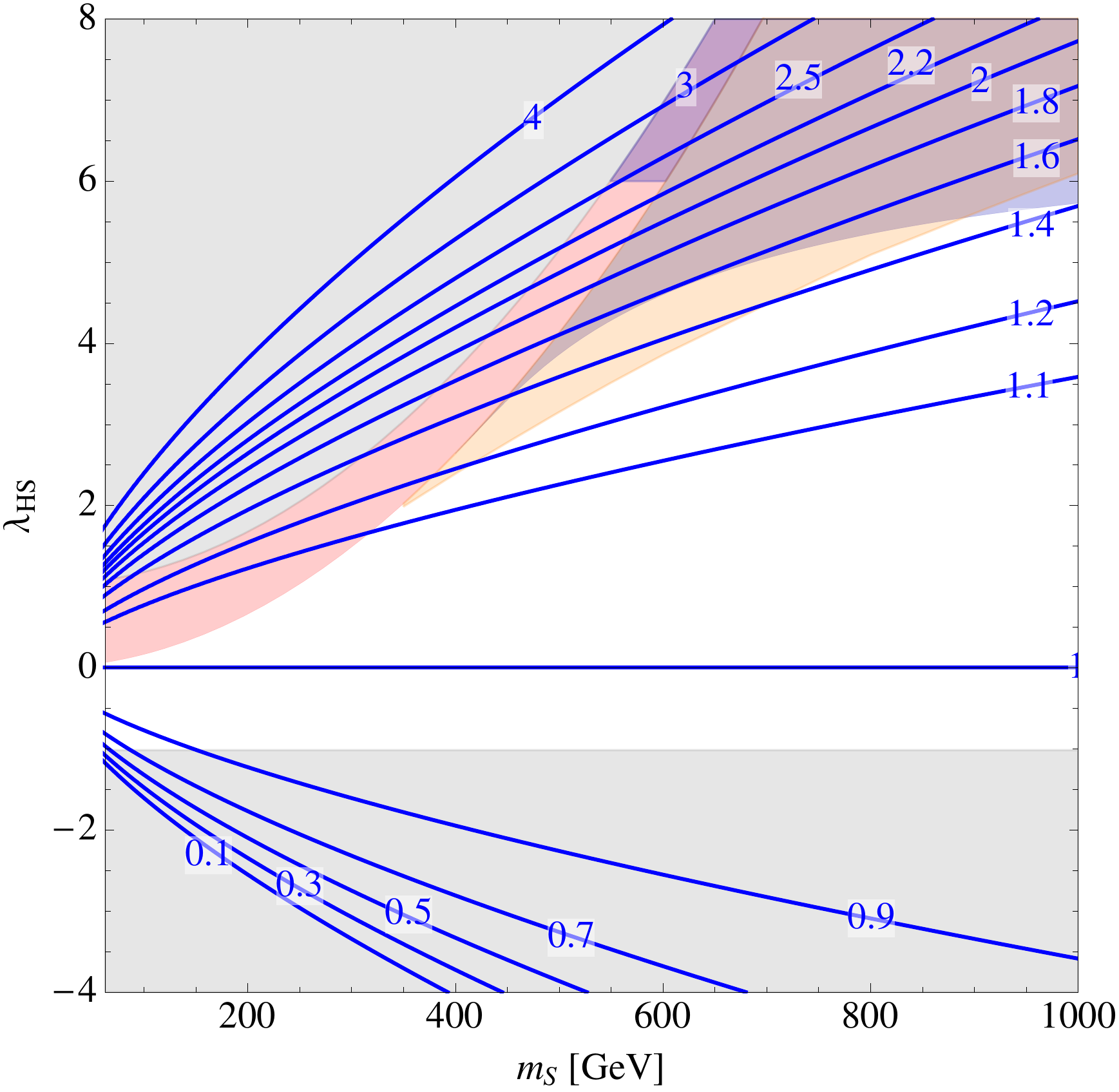}
\caption{Blue contours show $\lambda_3/\lambda_3^\mathrm{SM}$. Measuring $\lambda_3$ with a precision of 30\%, 20\%, and 8\% can be achieved at 14 TeV, 33 TeV, and 100 TeV hadron colliders with 3 ab$^{-1}$ of data, respectively. A 1000 GeV ILC with 2.5 ab$^{-1}$ could achieve a precision of 13\%. See text for details. }
\label{f.lambda3}
\end{figure}

One can measure $\lambda_3$ through double higgs production. The cross-section for producing a pair of higgs bosons is roughly three orders of magnitude smaller than the cross-section for producing a single higgs, which highlights the challenge of the measurement and the necessity for high luminosity.   Although the $4b$ final state has the largest rate, it also suffers from a huge QCD background. Instead, the most promising channel is in $bb\gamma\gamma$, whose main backgrounds are QCD and $t\bar t h$ production. Various studies have found that $\lambda_3$ can be measured between 30\%-50\% accuracy (at $1 \sigma$) at the 14 TeV LHC with 3 ab$^{-1}$ \cite{ATLAS:lambda3, Baglio:2012np, Goertz:2013kp, Barger:2013jfa,Yao:2013ika}. 
An early study for future hadron colliders \cite{Yao:2013ika} found that this accuracy can be refined to 20\% and 8\% for a 33 TeV and 100 TeV collider with $3 \iab$~\cite{Yao:2013ika} (at $1 \sigma$). More recent\footnote{This study appeared after the original publication of this paper.} work~\cite{Barr:2014sga}, which includes the effect of photon fakes, found that a 100 TeV collider with $3 \iab$ ($30\iab$) can achieve a $2 \sigma$ uncertainty of $\sim 20\%$ ($10\%$), with some dependence on background systematics.  Thus, taking $30\iab$ as a benchmark, we assume that a 100 TeV collider could exclude a $\lambda_3$ shift of about $10\%$.


The precision attainable for measuring $\lambda_3$ at lepton colliders is generally below that achievable at the HL-LHC. However, a high-luminosity, high-energy ILC with $\sqrt{s} = 1000 \gev$ and 2.5 ab$^{-1}$ of data could measure $\lambda_3$ with a precision of 13\%  \cite{Asner:2013psa, Tian:2013yda} (at $1 \sigma$).

The results of these studies imply that while it is unlikely a definitive exclusion will be achieved at a 14 or 33 TeV collider, a 100 TeV collider could exclude the entire one-step phase transition region of \fref{lambda3} (orange shaded region) with a confidence of better than 2 to 5 $\sigma$, depending on $m_S$ and luminosity. A high-energy ILC could exclude most, though not all, of the one-step transition region at the $2\sigma$ level.  Such measurements would also be sensitive to the two-step transition from tree-effects (red shaded region) for $\lambda_{HS} \gtrsim 1-2$.

\subsection{$Zh$ production cross section at Lepton colliders}
\label{ss.deltazh}

The singlet can also affect higgs couplings by generating a small correction to the higgs wave function renormalization, which modifies all higgs couplings by a potentially measurable amount. In particular, precision measurements of the $Zh$ production cross section at lepton colliders might be another avenue for indirect detection of such a singlet. \cite{Craig:2013xia}

At one loop, the fractional change in $Zh$ production relative to the SM prediction is given by~\cite{Englert:2013tya, Craig:2013xia}
\beq
\label{e.deltaZh}
\delta\sigma_{Zh}= \frac{1}{2}\frac{|\lambda_{HS}|^2 v^2}{4\pi^2 m_h^2}\left[1+F(\tau_\phi)\right]
\eeq
where we have modified the equation to comply with our convention of $v \approx 246 \gev$, substituted our coupling normalization $\lambda_i \to 2\lambda_{HS}$ and inserted a factor of $\frac{1}{2}$ since $S$ is a real and not a complex scalar. The loop function  $F(\tau)$, with $\tau_\phi=m_h^2/4m_S^2$, is given by
\beq
F(\tau)=\frac{1}{4\sqrt{\tau(\tau-1)}}\log\left(\frac{1-2\tau-2\sqrt{(\tau(\tau-1))}}{1-2\tau+2\sqrt{(\tau(\tau-1))}}\right).
\eeq

\begin{figure}
\centering
\includegraphics[width=8.5cm]{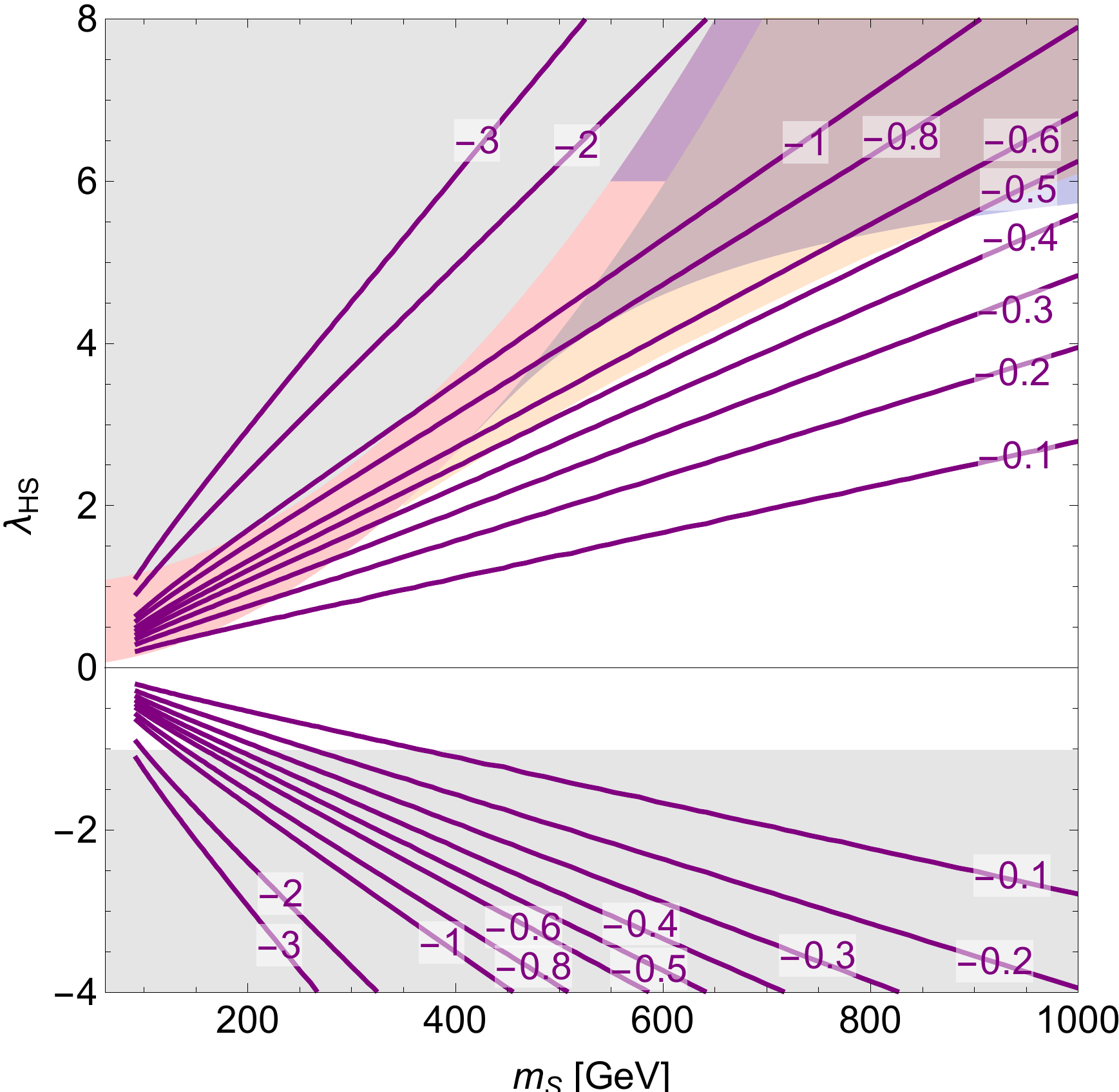}
\caption{Dashed blue contours: the one-loop corrections to the associated production cross-section of $Zh$ at lepton colliders \eref{deltaZh}, in \% relative to the SM.}
\label{f.deltaZH}
\vspace{1mm}
\end{figure}

%
%
%

$\delta \sigma_{Zh}$ is shown as a function of $(m_S, \lambda_{HS})$ in  \fref{deltaZH}. In the one-step region that is under perturbative control, $0.8\% \lesssim \delta \sigma_{Zh} \lesssim 0.5\%$. 
The proposed ILC, ILCLumiUp and TLEP programs could probe the $Zh$ cross section to about 2\%, 1\% and 0.3\% precision~\cite{Dawson:2013bba} respectively (at $1\sigma$). (See also \cite{Peskin:2012we,Blondel:2012ey,Klute:2013cx}.) The ILC therefore has limited utility to exclude EWBG in the nightmare scenario. However, TLEP could probe $\delta \sigma_{Zh} \approx 0.6\%$ at the $95\%$ confidence level. This is sensitive to a large fraction of the one- and two-step regions. 

It is useful to keep in mind that the precision of TLEP has a hard statistics limit \cite{Dawson:2013bba}. Without systematics, the $2\sigma$ precision of the $\sigma_{Zh}$ measurement with the data from 4 combined detectors is limited to 0.15\%, which could cover almost all of the EWBG-viable parameter space. 

It is clear that both indirect measurements, $\lambda_3$ at a 100 TeV collider and $\delta \sigma_{Zh}$ at TLEP, have great potential to detect the singlet-induced electroweak phase transition. These two measurements are in fact complementary, since they scale differently with $\lambda_{HS}$. This would allow the number of scalars running in the loops to be determined, a crucial detail of the theory. 

\section{Singlet Scalar Dark Matter}\label{s.dm}

We now consider the consequences of the singlet scalar $S$ acting as a stable thermal relic\footnote{A very similar computation was performed most recently in \cite{Cline:2013gha}, showing results in the same $(m_S, \lambda_{HS})$ plane as is relevant for our model. However, we repeat the calculation here for completeness, and to show how the resulting bounds overlap with the various regions in the nightmare scenario's parameter space.}.
This is not quite as unambiguous a consequence of EWBG as the bounds considered in Sections~\ref{s.collider} and \ref{s.indirect}. The hidden sector could be more complicated than just a singlet scalar, without the additional components affecting the phase transition. Indeed, we assume the presence of additional physics to generate the $CP$-violation necessary for EWBG. All of this could change the singlet scalar's cosmological history. Nevertheless, the minimal model could well be realized, and dark matter direct detection experiments represent a particularly exciting avenue for discovery in the relatively short term. 

The singlet thermal annihilation cross section has been presented in~\cite{Cline:2012hg}. Using standard methods~\cite{Kolb:1990vq}, it is straightforward to compute the relic density $\Omega_S$ and compare it to $\Omega_\mathrm{CDM} h_0^2 \approx 0.12$~\cite{Ade:2013zuv}, see \fref{singletdm} (left). As already pointed out e.g. in~\cite{Cline:2012hg}, in practically all of the parameter space relevant for EWBG, the large singlet-higgs coupling annihilates away much of the relic density, leaving the singlet scalar as a subdominant fraction of the dark matter density.

\begin{figure}
\begin{center}
\hspace*{-14mm}
\begin{tabular}{cc}
$\log_{10} \frac{\Omega_S}{\Omega_\mathrm{CDM}}$
&
$\log_{10} \left( \frac{\Omega_S}{\Omega\mathrm{CDM} } \ \times \ \sigma_S^\mathrm{SI} \right)$
\\
\includegraphics[width=8.5cm]{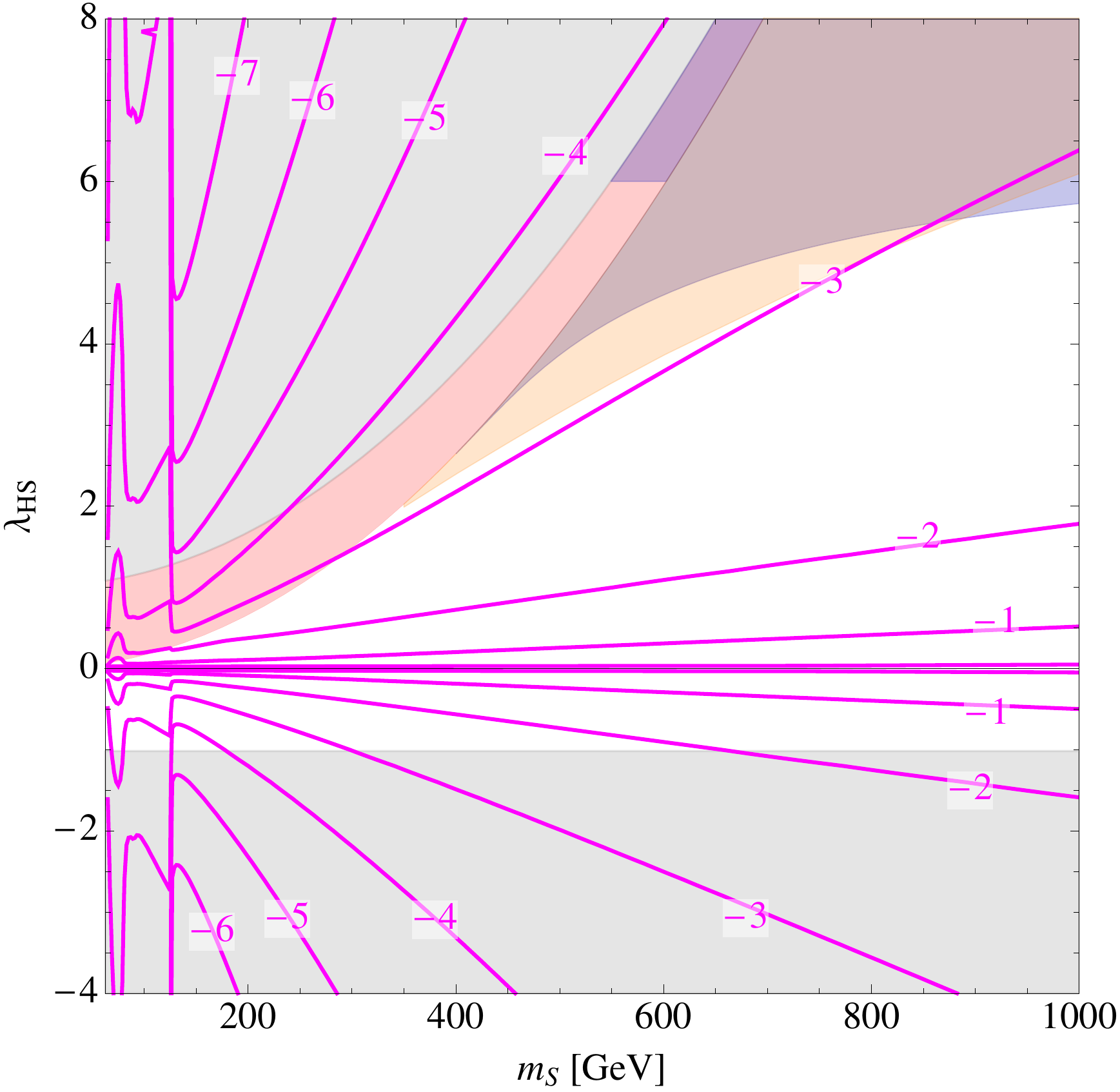}&
\includegraphics[width=8.5cm]{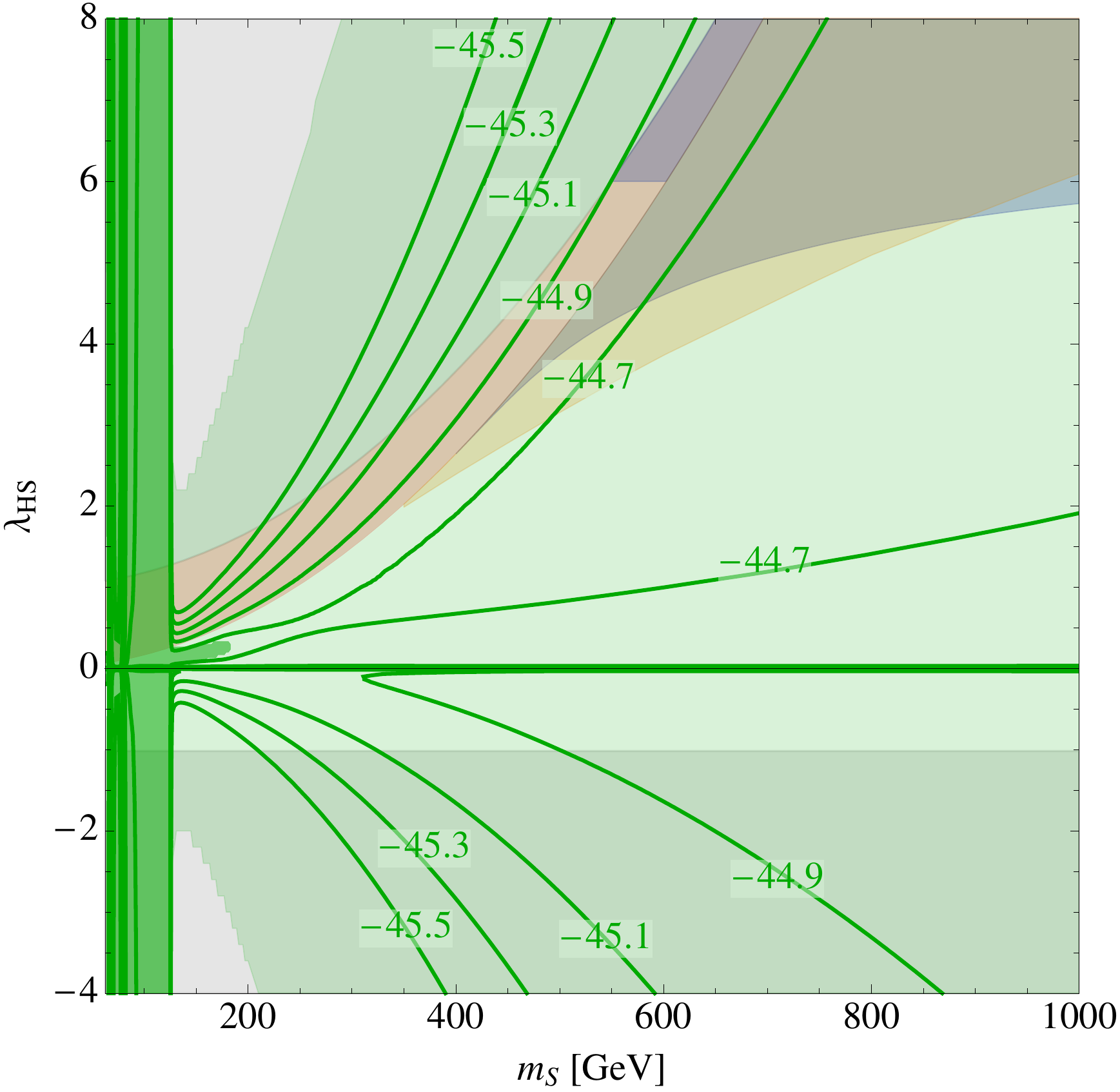}
\end{tabular}
\caption{
Dark matter properties of the singlet scalar S, assuming it is a stable thermal relic.
\textbf{Left}: magenta contours show contours of $\log_{10} \frac{\Omega_S}{\Omega_\mathrm{CDM}}$. In practically all of the parameter space viable for EWBG, the singlet scalar is a subdominant dark matter component. \textbf{Right}: green contours show the singlet scalar's direct detection cross section rescaled with relic density, $\log_{10} \left( \frac{\Omega_S}{\Omega\mathrm{CDM} } \ \times \ \sigma_S^\mathrm{SI} \right)$. The singlet-nucleon cross section is in units of $\mathrm{cm}^2$. The dark green shaded region is excluded by LUX \cite{Akerib:2013tjd}. The light green shaded region can be probed by XENON1T \cite{Aprile:2012zx}.
}
\label{f.singletdm}
\end{center}
\end{figure}

Direct detection constraints can be obtained by rescaling the cross section for higgs-mediated singlet-nucleon scattering \cite{Barbieri:2006dq, Cline:2012hg} by the relic density ratio $\Omega_S/\Omega_\mathrm{CDM}$. The resulting effective WIMP-nucleon scattering cross section is shown in \fref{singletdm} (right). The shaded dark green region is already excluded by LUX \cite{Akerib:2013tjd}, while the light-green region can be excluded at the future XENON1T experiment \cite{Aprile:2012zx}.

In these calculations, we have assumed the freeze-out temperature of $S$ to satisfy $T_f < T_c$. This is certainly true in the one-step region, where $T_c > 100 \gev$. In the two-step region, $T_f < 22 \gev$, and we find in  \sssref{twostep_full} that even $\lambda_S = \lambda_S^\mathrm{min} + 0.1$ results in $T_c = T_{c2} > 30 \gev$ for almost the entire two-step region. In this case, if $S$ is stable then the bounds calculated in this section apply to our model. However, it is possible to tune $\lambda_S \to \lambda_S^\mathrm{min}$ and achieve $T_c < T_f$. For this case, there are two possibilities for singlet freeze-out in the two-step phase transition region:
\begin{itemize}
\item The singlet freezes out in the unbroken phase at temperature $T_f^\mathrm{h=0}$. Since the universe resides in the singlet-VEV vacuum before the phase transition, the singlet can decay via $S \to hh$. This could deplete the singlet density to values much lower than indicated in \fref{singletdm} (left).
\item The singlet is in thermal equilibrium just before the phase transition at $T_c < 22 \gev$. If the singlet becomes lighter, it remains in thermal equilibrium and our above freeze-out estimate should apply. If it becomes heavier, it likely freezes out instantly.
\end{itemize}
Understanding the consequences of the second possibility would require further study, but it is clear that dark matter relic density may be considerably reduced in the two-step region, resulting in lower relic density and correspondingly weaker direct detection bounds than those shown in \fref{singletdm}.

That being said, assuming these direct detection bounds (with $T_f < T_c$ and a stable thermal relic $S$) apply to our model, the nightmare scenario for EWBG is already excluded for $m_S < m_h$ by LUX. Interestingly, the \emph{entire} EWBG-viable parameter space for both a one- and two-step phase transition is excludable at XENON1T. This provides a much earlier discovery possibility than a 100 TeV collider or a high-energy ILC.

\section{Strong Coupling Effects}\label{s.strongcouplings}

In large regions of our $(m_S, \lambda_{HS})$ parameter space we either manifestly have non-perturbative couplings, or relatively strong couplings to cause a one-loop first-order phase transition.  
In the non-perturbative regions, lack of theoretical control prevents us from conducting detailed studies. Nevertheless, we can make some qualitative statements about the possibility of a strong phase transition and its testability.  

There are two distinct regions with non-perturbative $\lambda_S$ in the  $(m_S, \lambda_{HS})$ plane. In the first, with negative $\mu_S^2$, the large $\lambda_S$ is required to ensure the EWSB vacuum is the universe's true ground state. In the second, with $\mu_S^2 > 0$ but $\lambda_{HS} < 0$, the large self-coupling is required to avoid a runaway potential for the singlet.  
In the absence of full theoretical control, the most conservative approach in examining these two regions is to assume that they maintain the basic vacuum structure implied by a naive classical analysis. 

Therefore, if the first region were viable, it would simple enlarge the allowed parameter space for a two-step phase transition in the direction of large $\lambda_{HS}$. These  strong phase transitions would be much more discoverable (by all experimental avenues) than the cases we have examined, meaning our statements about testability of the strong phase transition remain valid. 
The non-perturbative region with negative $\lambda_{HS}$ is more difficult to interpret. However, this region is not close to any region with a strong phase transition that is under theoretical control, and is likely not viable due to the appearance of a singlet runaway.

One may also ask whether there are any interesting effects due to large $\lambda_{HS}$ in regions where a strong phase transition is possible without large $\lambda_S$. Continuing our conservative line of reasoning, increasing $\lambda_{HS}$ would maintain the basic characteristics of the theory (strong phase transition) while making the theory even more testable. New phenomena may also arise in this direction, which has been considered previously in the context of a strongly interacting phase of the MSSM~\cite{Giudice:1998dj}.  In such a scenario, the singlets could turn into new composite states bound together by higgs exchange, similar to the stop-balls in~\cite{Giudice:1998dj}.

Finally, in regions of parameter space with moderately large but still perturbative couplings, understanding the theory's RG evolution is of critical importance. If the couplings required at low energy for a strong phase transition become non-perturbative at higher energies, it could invalidate our calculation of the universe's thermal history to find regions with acceptable phase transitions. 
The couplings need not stay perturbative to the GUT or Planck scale, merely in a sufficiently large range to ensure our calculations of the phase transition are trustworthy. To answer this question, we investigate the renormalization group equations (RGEs) of the nightmare scenario below.

\subsection{RG Evolution} 

The RGEs are easiest to work with in the \MSbar scheme. For completeness, we give the RGEs in this scheme in \aref{rges}.  
Note that this is a different choice than the on-shell scheme with cutoff regularization used in calculating the one-loop potential in \sref{ewpt}. 
These RGEs would naively suffice for understanding our model, since for small couplings, the physical matching calculation in the two schemes gives similar Lagrangian parameters. However, due to the large hierarchy of couplings $\lambda \ll \lambda_{HS}$, this correspondence breaks down in the one-step phase transition region. Therefore, we repeat some of our calculations in the \MSbar scheme, which we briefly summarize here. (This also serves as a useful cross-check.)

The zero-temperature one-loop correction to the higgs potential in \MSbar is given by
\begin{equation}
V_{0}^{CW, \overline{\mathrm{MS}}}(h)=\sum_i\frac{g_i(-1)^{F_i}}{64\pi^2}M_i^4\left(\log\frac{M_i^2}{\mu_r^2}+C_i\right)
\end{equation}
where the masses, $F_i$ and $g_i$ are the same as in \aref{FTEP}, $\mu_r$ is the renormalization scale, and $C_i = 5/6$ for vectors and $3/2$ for fermions and scalars. For a given choice of BSM parameters $(\mu^2_S, \lambda_{HS}, \lambda_S)$ we have to find $(\mu, \lambda)$ to set $m_h \approx 125 \gev$ and $v \approx 246 \gev$. We choose a renormalization scale of $\mu_r = 175 \gev$, and find that the required value for the Lagrangian parameter $\lambda$ is negative, though still $\mathcal{O}(0.1)$, when  $\lambda_{HS} \gtrsim 3$. This illustrates that negative quartic parameters do not necessarily signal a vacuum instability in the \MSbar scheme, since the resulting $V_\mathrm{eff}$ has no runaways with arbitrarily high field values consistent with our perturbative expansion. 

In the on-shell scheme, $m_S$ and $\lambda_{HS}$ correspond to the physical observables of mass and $hSS$ coupling respectively. This is not  true in \MSbar, but the ``effective" $\lambda_{HS}$ coupling 
\begin{equation}
\lambda_{HS}^\mathrm{eff} \equiv \frac{1}{2v} \left. \frac{ \partial^3 V_\mathrm{eff}}{\partial^2 S \partial h} \right|_{h=v, S=0} = \lambda_{HS} \left[ 1 + \frac{3}{16 \pi^2} \left( \lambda \log \frac{3 \lambda v^2 - \mu^2}{\mu_r^2} + \lambda_S \log \frac{\mu_S^2 + \lambda_{HS} v^2}{\mu_r^2} \right) \right]
\end{equation}
is within a few (ten) percent of $\lambda_{HS}$ for $\lambda_S \sim 0.1 (1)$.  Thus, the  $(m_S, \lambda_{HS})$ plane  in \MSbar is approximately equivalent to the same plane in on-shell. 

Finally, we compare the shape of the zero-temperature effective potential in this plane obtained using \MSbar and on-shell. On the lower boundary of the one-step phase transition region they are nearly indistinguishable, while some scheme-dependent differences become apparent as we raise $\lambda_{HS}$ to values where the perturbative expansion is untrustworthy according to \sssref{perturbativereliability}.\footnote{We find $(v_c/T_c)_{\overline{\mathrm{MS}}} \geq (v_c/T_c)_{\mathrm{on-shell}}$.} Since the important $W, Z, t$ contributions to the thermal potential are the same in both schemes, our determination of the one-step phase transition region is robust across different scheme choices. \footnote{The tree-level argument leading to the derivation of the two-step phase transition region in \ssref{twostep} are unchanged.}

We are now ready to examine the RG evolution of the model. The boundary conditions are set at $\mu_r = \mu_r^0 =  175 \gev$. Fixing $\lambda_S(\mu_r^0) = 0$ and setting $\lambda(\mu_r^0)$ to the value obtained by the physical matching calculation, we find that the theory remains perturbative up to scales of $\sim 10 \ (100) \tev$ for $\lambda_{HS}(\mu_r^0) \lesssim 3 \ (4)$. This conclusion is not significantly altered by letting $|\lambda_S(\mu_r^0)| \sim \mathcal{O}(1)$. 

Therefore our analysis of the phase transition is sound in most of the region where we claimed perturbative reliability in \sssref{perturbativereliability}. Furthermore, additional hidden-sector physics must enter before the 10-100 TeV scale if the one-step phase transition is realized, but this does not influence our calculation of the phase transition. 

Requiring no Landau Poles up to $\sim 100 \tev$ could also slightly expand our definition of the non-perturbative (gray) regions in the $(m_S, \lambda_{HS})$ plane, but this does not affect our conclusions regarding the detectability of the phase transition.

We conclude our RG discussion with a final comment on vacuum stability. 
It is well understood that for the measured higgs mass in the SM, the universe is in a metastable state \cite{Buttazzo:2013uya}, since the $y_t^4$ term in the $\lambda$ RGE pushes the quartic down towards negative values at high energies. 
\eref{RGE} makes clear that this can be counteracted by turning on a positive $\lambda_{HS}$ coupling, where $\lambda_{HS} \lesssim 1$ to avoid $\lambda_S$ becoming non-perturbative before the GUT scale. Therefore, there exists a part of the viable EWBG parameter space in the two-step region near $(m_S, \lambda_{HS}) \sim (200 \gev, 0.5)$ that is valid to high scales, and also allows for an absolutely stable universe.  
Interestingly, this is in the most difficult part of the EWBG-viable region to test, with small couplings that will require the highest energy and luminosity to investigate.

\section{Conclusions}\label{s.discussion}

\begin{figure}
\centering
\includegraphics[width=12cm]{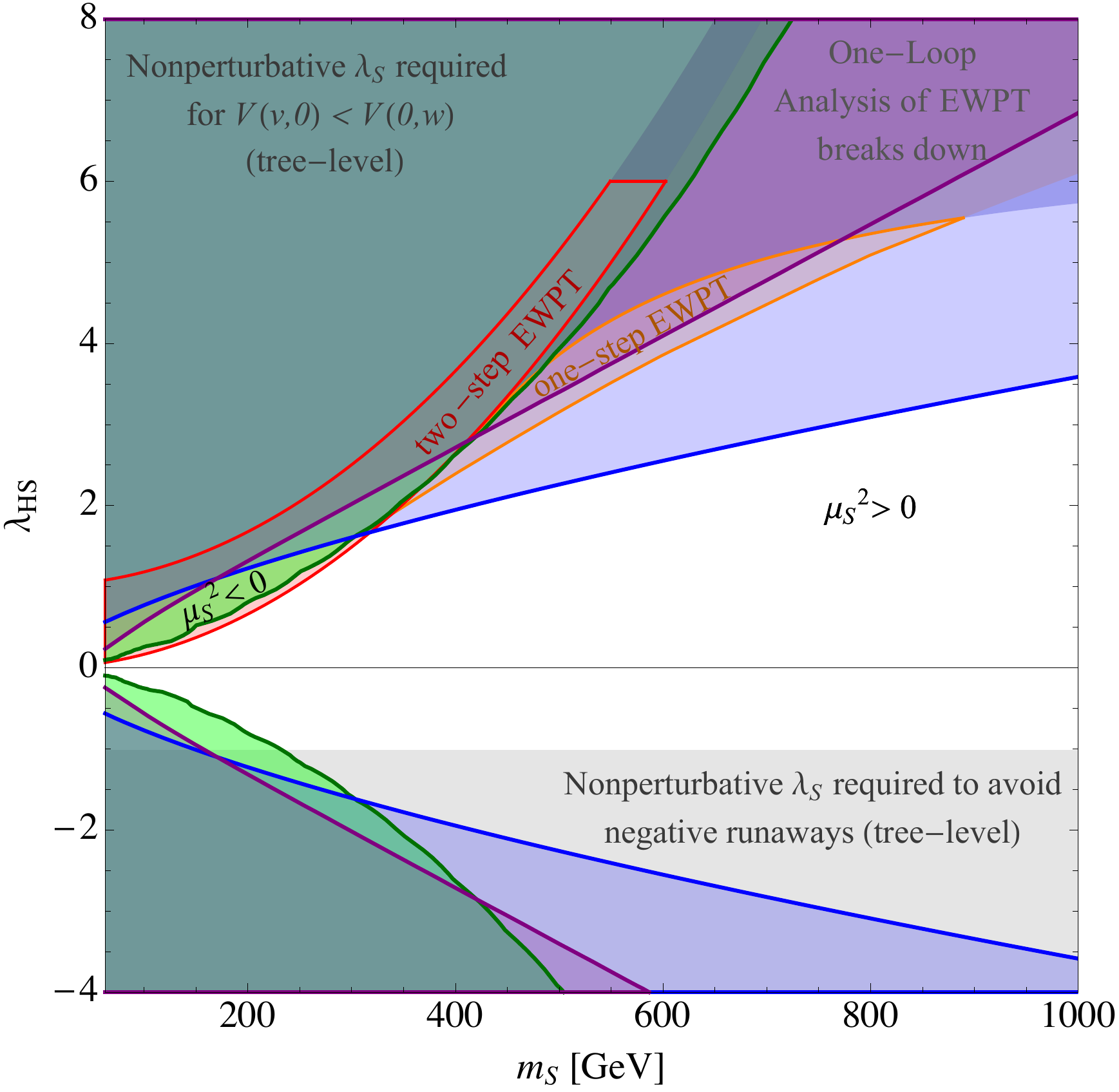}
\caption{
Summary of the nightmare scenario's parameter space. \textbf{Gray} shaded regions require non-perturbative $\lambda_S > 8$ and are not under theoretical control, see \ssref{couplingplane}. \textbf{Red} shaded region with red boundary: a strong two-step PT from tree-effects is possible for some choice of $\lambda_S$, see \ssref{onestep}. \textbf{Orange} shaded region with orange boundary: a strong one-step PT from zero-temperature loop-effects is possible, see \sssref{ewptvialoop}. Gray-Blue shading in top-right corner indicates the one-loop analysis becomes unreliable for $\lambda_{HS} \gtrsim 5 (6)$ in the one-step (two-step) region, see \sssref{perturbativereliability} and \ref{sss.twostep_full}.
In the \textbf{blue} shaded region,
higgs triple coupling is modified by more than 10\% compared to the SM, which could be excluded at the $2\sigma$ level by a 100 TeV collider, see \ssref{triplehiggs}. 
In the \textbf{green} shaded region, our simple collider analysis yields $S/\sqrt{B} \leq 2$
for VBF production of $h^*\to S S$  at a 100 TeV collider, see \sref{collider}. (In both cases assume 30 $\mathrm{ab}^{-1}$ of data.) 
In the \textbf{purple} shaded region, $\delta \sigma_{Zh}$ is shifted by more than 0.6\%, which can be excluded by TLEP, see \ssref{deltazh}.
Note that both EWBG preferred regions are excludable by XENON1T if $S$ is a thermal relic, see \sref{dm}.
}
\label{f.summaryplot}
\end{figure}

In this paper, we have started investigating the possibility of formulating a ``no-lose'' theorem for testing EWBG at future colliders. To this end, we consider a ``nightmare scenario'' which minimizes experimental testability while realizing several different mechanisms of generating a strong first-order EW phase transition. 

The nightmare scenario is simple -- we add one real scalar singlet to the SM, which couples through the higgs portal with a $\mathbb Z_2$ symmetry and has a mass of greater than half the higgs mass. The entire parameter space of this nightmare scenario can be represented in the $(m_S,\lambda_{HS})$ plane and we provide a diagrammatic summary of our findings in \fref{summaryplot}. There are two distinct regions allowing for a strong electroweak phase transition, a one-step transition marked in orange and a two-step (or thermal) transition marked in red. The blue region marks where indirect measurements of the triple-higgs coupling $\lambda_3$ at a 100 TeV collider are sensitive, while the green region marks where direct searches through VBF production of $h^*\to SS$ at a 100 TeV collider are sensitive. The purple region shows where TLEP can probe the scenario by measuring $\delta \sigma_{Zh}$.

The entire one-step phase transition region, and much of the two-step region, can be probed with the $\lambda_3$ and $\delta \sigma_{Zh}$ measurements. Furthermore, our simple collider analysis for the sensitivity of VBF direct singlet production yields $S/\sqrt{B} > 2$ in almost the entire two-step region. 
It may therefore be possible to exclude the entire two-step region with a more complete analysis \cite{NateMattArun}, or with more optimistic assumptions for the capabilities of a future 100 TeV collider.


The experimental outlook for probing electroweak baryogenesis is, in fact, very positive. As outlined in the introduction, non-minimal models offer many discovery avenues. The nightmare scenario we study in this paper was selected for its minimal experimental signatures, but it too cannot escape detection completely.  While proposed linear colliders have limited utility in this case, both future circular lepton colliders (like TLEP) and a future 100 TeV pp collider (like SPPC/FCC) have great sensitivity:
\begin{itemize}
\item The triple higgs coupling and direct higgs portal production measurements at a 100 TeV collider with $30 \iab$ together cover practically the entire parameter space which leads to a strong singlet-induced electroweak phase transition in the nightmare scenario. Reaching this sensitivity goal should serve as a useful design benchmark of a future 100 TeV collider program.
\item The TLEP measurement of $\delta \sigma_{Zh}$ is sensitive to a significant fraction of the one- and two-step phase transition regions. A 1000 GeV ILC with 2.5 ab$^{-1}$ could also exclude triple higgs coupling deviations larger than $\approx 25\%$ \cite{Asner:2013psa, Tian:2013yda}, which covers a similar part of parameter space. 
\end{itemize}
Furthermore, the measurements possible at the two kinds of colliders are complementary, allowing details of the theory (like number of new scalar fields) to be determined.

Future dark matter searches have the potential to beat future colliders to the punch in observing low-lying EW states. This was already observed in \cite{Low:2014cba} for neutralino dark matter. %
In our scenario, if the scalar $S$ is a thermal relic with $T_f < T_c$, then the \emph{entire} EWBG-viable parameter space of the nightmare scenario can be ruled out by XENON1T. However, as with all DM related searches, this exclusion depends on the cosmological history and, as mentioned in \sref{dm}, could be altered in our scenario without influencing the phase transition.

Our study yields several avenues for future investigation. 
Given that much or all of our nightmare scenario's parameter space is in reach of either 100 TeV collider, the question arises whether a more rigorous ``no-lose" theorem for EWBG at future colliders can be constructed. 
The nightmare scenario could in principle be made even more difficult to discover. Thus, it would be interesting to explore the extent to which its experimental signatures can be suppressed while maintaining a strong phase transition. 
There may be scenarios in which \emph{exclusion} has to proceed by investigating the required new sources of CP violation. 
Regardless of the model, if EWBG is realized in our universe, \emph{confirming} this will require studying the phase transition together with the new sources of CP violation.

\bigskip

{\em Acknowledgements:} 
The authors are especially grateful to Nathaniel Craig for prompting us to improve the collider analysis in an earlier version of this paper, and for fruitful subsequent discussions. 
The authors wish to thank Nima Arkani-Hamed, Daniel Chung, Tao Han, Roni Harnik, Stephen Martin, David Morissey, Mariano Quiros, Matthew Reece, and Lian-Tao Wang for useful discussions and comments. The work of D.C. and C.-T.Y. is supported in part by the National Science Foundation under Grant PHY-PHY-0969739. The work of P.M. is supported in part by NSF CAREER Award NSF-PHY-1056833. Part of this work was completed at the Aspen Center for Physics, which operates under the NSF Grant 1066293.

\appendix
\section{Finite-Temperature Effective Potential}\label{a.FTEP}
The four components of the finite-temperature effective potential 
\begin{equation}
V_\mathrm{eff}(h,T)=V_0(h)+V_0^{CW}(h)+V_T(h,T)+V_r(h,T).
\end{equation}
are given as follows. 
$V_0$ is the tree-level potential defined in \eref{V0}. The Coleman-Weinberg potential $V_0^{CW}$ \cite{Coleman:1973jx} is the zero-temperature one-loop correction. In the on-shell renormalization scheme with cutoff regularization it is given by 
\begin{equation}
\label{e.Vcw}
 V_0^{CW} = \sum_i (-1)^{F_i} \ \frac{g_i}{64 \pi^2} \left[ m_i^4(h) \left( \log \frac{m_i^2(h)}{m_i^2(v)}  - \frac{3}{2}\right) + 2 m_i^2(h) m_i^2(v)\right],
\end{equation}
where we have applied the renormalization conditions 
\begin{equation}
\left. {V_0^{CW}}'(h)\right|_{h=v} = 0 \ \ , \ \ \ \  \left. {V_0^{CW}}''(h)\right|_{h=v} = 0
\end{equation}
so the higgs mass and VEV are not perturbed from their tree-level values, see e.g. \cite{Delaunay:2007wb, Quiros:1999jp}. These renormalization conditions also ensure that the $hSS$ coupling is not modified from its tree-level value of $-2 v \lambda_{HS}$ at one-loop, naturally allowing us to phrase our results in terms of physical parameters. $F_i$ is fermion number, 1 for fermions and 0 for bosons. Following the notation of \cite{Noble:2007kk}, the masses of the SM and BSM particles are given by $M_i^2(h)=M_{0,i}^2+a_i h^2$, where the relevant contributions are
\barray
i&=&(t,W,Z,h,G,S)\nonumber\\
M_{0,i}^2&=&(0,0,0,-\mu^2,-\mu^2,\mu_S^2)\nonumber\\
a_i&=&\left(\frac{\lambda_t^2}{2},\frac{g^2}{4},\frac{g^2+g'^2}{4},3\lambda,\lambda,\lambda_{HS}\right)\\
g_i&=&\left(
12, 6, 3, 1, 1, 1
\right).
\earray
In practice we neglect the numerically insignificant Goldstone contributions as well, since handling them correctly near $h = v$ takes special care \cite{Martin:2014bca}.

The one-loop finite temperature potential is given by \cite{Dolan:1973qd,Weinberg:1974hy}
\begin{equation}
V_T(h,T) =\sum_i (-1)^{F_i} \  \frac{g_i T}{2\pi^2}\int dk k^2\log\left[1- (-1)^{F_i}\exp\left(\frac{1}{T}\sqrt{k^2+M_i^2(h)}\right)\right]
\end{equation}
For $T\gg M_i$, the boson thermal contributions contain multi-loop infrared-divergences which must be resummed by adding ring terms,
\beq
V_r(h,T)=\sum_i\frac{T}{12\pi}\Tr\left[M_i^3(h)-(M_i^2(h)+\Pi_i(0))^{3/2}\right],
\eeq
where $i$ runs over the light bosonic degrees of freedom and $\Pi_i(0)$ is the zero-momentum polarization tensor \cite{Comelli:1996vm}:
\begin{eqnarray}
\Pi_h(0) &=& \Pi_G(0) = T^2 \left(\frac{3}{16} g^2 + \frac{1}{16} {g'}^2 + \frac{1}{4} \lambda_t^2 + \frac{1}{2} \lambda + \frac{1}{12} \lambda_{HS} \right) \ ,
\\
\nonumber\Pi_S(0) &=& T^2 \left( \frac{1}{3}  \lambda_{HS} + \frac{1}{4} \lambda_{S}\right)\ ,
\\
\nonumber \Pi_{GB}(0) &=& \frac{11}{6} T^2  \ \mathrm{diag}(g^2, g^2, g^2, {g'}^2) \ ,
\end{eqnarray} 
and we use the gauge boson mass matrix in gauge basis
\begin{equation}
M^2_{GB}(h) = \frac{h^2}{4} \left(
\begin{array}{cccc}
g^2 & 0 & 0 & 0\\
0 & g^2 & 0 & 0\\
0 & 0 & g^2 & -g g'\\
0 & 0 & -g g' & {g'}^2
\end{array}
\right) \ .
\end{equation}

\section{Renormalization Group Equations}
\label{a.rges}

The one-loop RGEs in the \MSbar scheme are 

\begin{eqnarray}
\nonumber 16\pi^2 \frac{d\lambda}{dt}&=&\frac{3}{8}g_1^4+\frac{9}{8}g_2^4+\frac{3}{4} g_1^2 g_2^2 - 6 y_t^4 + 24 \lambda^2 + 12 y_t^2\lambda - 3 g_1^2 \lambda -9g_2^2\lambda+\frac{1}{2}\lambda_{HS}^2\\
\nonumber 16\pi^2\frac{d\lambda_{HS}}{dt}&=&\lambda_{HS}(12\lambda+6\lambda_S+4\lambda_{HS}+6 y_t^2-\frac{3}{2}g_1^2-\frac{9}{2}g_2^2)\\
\nonumber  16\pi^2\frac{d\lambda_S}{dt}&=&2\lambda_{HS}^2+18\lambda_S^2\\
\label{e.RGE} 16\pi^2\frac{d g_1}{dt}&=&\frac{41}{6}g_1^3\\
\nonumber  16\pi^2\frac{d g_2}{dt}&=&-\frac{19}{6}g_2^3\\
\nonumber  16\pi^2\frac{d g_3}{dt}&=&-7g_3^3\\
\nonumber  16\pi^2\frac{d y_t}{dt}&=&y_t\left(\frac{9}{2}y_t^2-\frac{17}{12}g_1^2-\frac{9}{4}g_2^2-8g_3^2\right).
\label{eq:RGEs}
\end{eqnarray}

\bibliography{SMSewbg}
\bibliographystyle{JHEP}


\end{document}